\documentclass[twoside]{article}
\usepackage{epstopdf}
\usepackage{amsfonts,amssymb,amsbsy,textcomp,marvosym,picins,amsmath,caption,threeparttable,amsthm,subfigure}
\usepackage{eurosym,mathrsfs,fancyhdr,CJK,multicol,graphics,indentfirst,color,bm,upgreek,booktabs,graphicx,float}
\usepackage{multirow} 
\usepackage{xcolor}  
\usepackage{subfig}
\usepackage{multirow}
\usepackage{amssymb} 
\usepackage{booktabs}
\usepackage{tabularx}
\usepackage{setspace}
\newcommand{\cl}[1]{{\color{black}#1}}
\newcommand{\cll}[1]{{\color{black}#1}}
\newcommand{\clr}[1]{{\color{black}#1}}

\newcommand{\etal}{\textit{et al.}}

\def\footnoterule{\kern 1mm \hrule width 10cm \kern 2mm}

\allowdisplaybreaks
\catcode`@=11
\def\title#1{\vspace{3mm}\begin{flushleft}\vglue-.1cm\Large\bf\boldmath\protect\baselineskip=18pt plus.2pt minus.1pt #1
\end{flushleft}\vspace{1mm} }
\def\author#1{\begin{flushleft}\normalsize #1\end{flushleft}\vspace*{-4pt} \vspace{3mm}}
\def\address#1#2{\begin{flushleft}\vglue-.35cm${}^{#1}$\small\it #2\vglue-.35cm\end{flushleft}\vspace{-2mm}\par}

\catcode`@=11
\def\section{\@startsection{section}{1}{\z@}%
 {-3ex \@plus -.3ex \@minus -.2ex}%
 {2.2ex \@plus.2ex}%
{\normalfont\normalsize\protect\baselineskip=14.5pt plus.2pt minus.2pt\bfseries}}
\def\subsection{\@startsection{subsection}{2}{\z@}%
 {-3ex\@plus -.2ex \@minus -.2ex}%
 {2ex \@plus.2ex}%
{\normalfont\normalsize\protect\baselineskip=12.5pt plus.2pt minus.2pt\bfseries}}
\def\subsubsection{\@startsection{subsubsection}{3}{\z@}%
 {-2.2ex\@plus -.21ex \@minus -.2ex}%
 {1.4ex \@plus.2ex}
{\normalfont\normalsize\protect\baselineskip=12pt plus.2pt minus.2pt\sl}}

\begin{document}
\begin{CJK*}{GBK}{song}
\thispagestyle{empty}
\vspace*{-13mm}
\noindent {\small Chen L, Ye JT, Zhang XP. Multi-feature super-resolution network for cloth wrinkle synthesis.
JOURNAL OF COMPUTER SCIENCE AND TECHNOLOGY \ 36(3): 478 - 493
\ May 2021. DOI 10.1007/s11390-021-1331-y}
\vspace*{2mm}

\title{Multi-Feature Super-Resolution Network for Cloth Wrinkle Synthesis}

\clr{\author{Lan Chen$^{1,2}$, {\itshape Member, CCF}, Juntao Ye$^{1}$, {\itshape Member, CCF} and Xiaopeng Zhang$^{1,3,*}$, {\itshape Member, CCF, ACM, IEEE}}}

\address{1}{\clr{National Laboratory of Pattern Recognition, Institute of Automation, Chinese Academy of Sciences, Beijing, China}}
\address{2}{\clr{The School of Artificial Intelligence, University of Chinese Academy of Sciences, Beijing, China}}
\address{3}{\clr{Zhejiang Lab, Hangzhou, Zhejiang Province, China}}

\vspace{2mm}

\noindent E-mail: chenlan2016@ia.ac.cn; yejuntao@gmail.com; xiaopeng.zhang@ia.ac.cn   \\[-1mm]

\noindent Received January 29, 2021; accepted April 27, 2021.\\[1mm]

\let\thefootnote\relax\footnotetext{{}\\[-4mm]\indent\ Regular Paper\\[.5mm]
\indent\quad Special Section of CVM 2021\\[.5mm]
\indent\ \clr{This work is supported by the National Key Research and Development Program of China under Grant No. 2018YFB2100602, the National Natural Science Foundation of China under Grants Nos. 61972459, 61971418 and 62071157, and Open Research Projects of Zhejiang Lab under Grant No. 2021KE0AB07.}\\[.5mm]
\indent\ $^*$Corresponding Author
\\[.5mm]\indent\ \copyright Institute of Computing Technology, Chinese Academy of Sciences 2021}

\noindent {\small\bf Abstract} \quad  {\small {Existing physical cloth simulators suffer from expensive computation and difficulties in tuning mechanical parameters to get desired wrinkling behaviors.
	Data-driven methods provide an alternative solution. \clr{They typically synthesize cloth animation at a much lower computational cost, and also create wrinkling effects that are similar to the training data. }
	In this paper we propose a deep learning based method for synthesizing cloth animation with high resolution meshes. 
	To do this we first create a dataset for training: a pair of low and high resolution meshes are simulated and their motions are synchronized.
	As a result the two meshes exhibit similar large-scale deformation but different small wrinkles. 
	Each simulated mesh pair \clr{is} then converted into a pair of low and high resolution \cl{``images''} 
	(a 2D array of samples), with each image pixel being interpreted as any of three descriptors:
	the displacement, the normal and the velocity.
	With these image pairs, we design a multi-feature super-resolution (MFSR) network that jointly \clr{trains} an upsampling synthesizer for the three descriptors.
	The MFSR architecture consists of shared and task-specific layers to learn multi-level features when super-resolving three descriptors simultaneously.
	Frame-to-frame consistency is well maintained thanks to the proposed kinematics-based loss function. 
	Our method achieves realistic results at high frame rates: $12 \sim 14$ times faster than traditional physical simulation.
	We demonstrate the performance of our method with various experimental scenes, including a dressed character with sophisticated collisions.}}

\vspace*{3mm}

\noindent{\small\bf Keywords} \quad {\small cloth animation, deep learning, multi-feature, super-resolution, wrinkle synthesis}

\vspace*{4mm}

\end{CJK*}
\baselineskip=18pt plus.2pt minus.2pt
\parskip=0pt plus.2pt minus0.2pt
\begin{multicols}{2}

\section{Introduction}

Cloth animation plays an important role in many  applications, such as movies, video games \clr{and} virtual try-on \cl{\cite{liang2020machine, wang2020vr}}. 
With the rapid development of physics-based simulation techniques \cite{terzopoulos87elastically,provot97collision,BW98,bridson03wrinkles}, garment animations with remarkably realistic and detailed folding patterns can be achieved.
However, these techniques require \clr{high-resolution} meshes to represent fine details, \clr{and} therefore need much computation to solve velocity-updating equations and resolve collisions. Moreover it is labor-intensive to tune simulation parameters for a desired wrinkling behavior. 
Recently data-driven methods \cite{wang10example,zurdo2013wrinkles,santesteban2019learning} provide alternative solutions for these problems, as they offer fast production and also create wrinkling effects that highly resemble the training data.
Relying on precomputed data and data-driven techniques, a high-resolution (HR) mesh is either directly synthesized, or super-resolved from a physically simulated low-resolution (LR) mesh.  
Nevertheless, existing data-driven methods either depend on human body poses \cite{wang10example,santesteban2019learning,Feng2010transfer,deAguiar10Stable} \clr{and} thus are not suitable for loose garments, or lack of dynamic modeling of wrinkle behaviors \cite{zurdo2013wrinkles,kavan11physics,chen2018synthesizing,oh2018hierarchical,laehner2018deepwrinkles} for general case of free-flowing cloth.

To tackle these challenges, we propose a framework, synthesizing cloth wrinkles with a deep learning based method.
We create datasets, from physics-based simulation, as the training data. 
The simulation is assumed to be independent of human bodies and not limited to tight garments.
This dataset is generated by a pair of LR and HR meshes with synchronized simulations.
Given the simulated mesh pairs, we aim to map the LR meshes to the HR domain by a detail enhancement method, which is essentially a super-resolution (SR) operation.
Deep SR networks have proven to be powerful and fast machine learning tools for image detail enhancement \cite{ledig2017photo, zhang2018residual}.
Yet for surface meshes which usually have irregular structures, it is not straightforward to apply traditional convolutional operations as for images.
Chen \etal \cite{chen2018synthesizing} proposed a method, converting manifold meshes into geometry images \cite{gu2002geometry}, to solve this issue.
Inspired by their work, we design a multi-feature super-resolution network (MFSR) to improve the synthesized results and model the dynamic wrinkle behaviors.
The LR and HR image pairs, encoding three features: the displacement, the normal and the velocity, are fed into the network for training. 
Our MFSR jointly learns upsampling synthesizers with a multi-task architecture, consisting of a shared network and three task-specific networks, instead of combining all features with a single SR network.
The proposed spatial and temporal losses also contribute to the generation of dynamic wrinkles and further maintain frame-to-frame consistency.
At runtime, with super-resolved geometry images generated by MFSR, we convert them back into HR meshes.  
As our approach is based on deep neural networks, it reduces the computational cost significantly.
In summary, the main contributions of our work are as \clr{follows.}
\begin{itemize}
	\item We propose a novel framework for cloth wrinkle synthesis, which is composed of synchronized simulation, mesh-image conversion, and a multi-feature super-resolution network (MFSR). 
	\item We learn both shared and task-specific representations of garment shapes via multiple features.
	\item We generate dynamic wrinkles and consistent mesh sequences thanks to the spatial and temporal loss functions.
\end{itemize}

We qualitatively and quantitatively evaluate our method for
various cloth types (tablecloths and long skirts) and motion sequences. 
Experimental results show that the quality of synthesized garments is comparable with that from a physics-based simulation, yet significantly reducing the computation cost.
To the best of our knowledge, this is the first approach to employ a multi-feature learning model on 3D dynamic wrinkle synthesis.

 
\section{\clr{Related Work}}

\subsection{\clr{Cloth Animation}}
A physics-based simulation for realistic fabrics includes velocity updating by physical energies \cite{terzopoulos87elastically,bridson03wrinkles}, time integration \cite{BW98}, collision detection and collision response \cite{provot97collision}.
These modules are solved separately and \clr{time-consuming}.
To improve the efficiency of this system, researchers have exploited many algorithms such as implicit time integration \cite{BW98}, adaptive remeshing \cite{Narain2012AAR} and iterative optimization \cite{liu13fast}.
Nevertheless, these algorithms still cost the expensive computation to produce rich wrinkles and are \clr{labor-consuming} to tune mechanical parameters for desired wrinkling behaviors.
Recently data-driven methods have drawn much attention as they offer faster cloth animations than the physics-based methods.
Based on precomputed data and data-driven techniques, an HR mesh is either directly synthesized, or super-resolved from a physically simulated LR mesh.
In the first stream of work, with precomputed data, researchers have investigated many techniques to accelerate the process for new animations, 
such as a linear conditional model \cite{deAguiar10Stable, Guan12DRAPE} and a secondary motion graph \cite{Kim2013near}.
Additionally, deep learning-based methods 
\cite{gundogdu2019garnet, Wang2018garmentdesign, wang2019learning} are also used to generate garments on human bodies.   
In the another line of work, researchers have proposed to combine coarse mesh simulations with learned geometric details from paired mesh databases, to generalize the performance to complicated testing scenes.
This stream of methods includes wrinkle synthesis depending on bone clusters \cite{Feng2010transfer} or human poses \cite{wang10example} for fitted clothes, and linear upsampling operators \cite{kavan11physics} or low-dimensional subspace with bases \cite{zurdo2013wrinkles, Hahn2014subspace} for general case of free-flowing cloth.
Inspired by these data-driven methods, we propose a deep learning based approach to synthesize wrinkles on coarse simulated meshes, while our approach is independent with poses or skeletons and not limited with tight garments.
 
\subsection{Representation in 3D Learning}
To process 3D models for deep learning, there are various representations \cite{xiao2020survey,yuan2021revisit}, \clr{e.g.}, voxels, images, point clouds, meshes.
Wang \etal \cite{wang2017cnn} \clr{used} voxel grids with octree-based convolutional neural networks (CNNs) for 3D shape analysis.
Su \etal \cite{Su2015mvcnn} \clr{learned} to recognize 3D shapes from multi-view images with 2D-CNNs.
Representations based on voxels or multi-view images are extrinsic to the shapes, which are sensitive to   isometric deformations, like rotation or translation. 
Instead of rendered images, recent works \cite{chen2018synthesizing,sinha2016deep} use a technique called geometry images \cite{gu2002geometry} encoding features of 3D meshes into \clr{the} 2D domain for 3D object recognition and generation .
With a patch-based approach, this technique is easily coped with deep CNNs \clr{and} thus suitable for our mesh super-resolution task.
Geometry images require parameterization for non-rectangular meshes, \clr{and} we use a padding scheme to avoid mesh distortion.
Recently, some researches \cite{tan2018autoencoder, tan2018variational} directly encode triangle meshes with deformation-based features \cite{gao2016efficient, gao2019sparse} into latent space with applications to shape embedding and synthesis.
These methods focus on the deformation of overall \clr{meshes. However, our} patch-based algorithm aims at learning local details and is independent of the underlying mesh connectivity.  
  
Feature-based methods aim for proper descriptions of irregular 3D meshes, for synthesizing detailed and realistic objects.
Conventional data-driven methods \cite{zurdo2013wrinkles} simplify the calculation of wrinkle features, by formulating the strain or stress in an LR mesh.
As for deep learning, several algorithms have also investigated robust descriptors for wrinkle deformation. 
Chen \etal \cite{chen2018synthesizing} and Oh \etal \cite{oh2018hierarchical} used 3D coordinates to augment coarse meshes with synthesized wrinkles.
Wang \etal \cite{wang2019learning} learned an autoencoder network for cloth using 3D positions. 
Instead, Santesteban \etal \cite{santesteban2019learning} decomposed the cloth deformation into two displacements, a global fit displacement and the wrinkle displacements.
In addition to the position or the displacement, L\" ahner \etal \cite{laehner2018deepwrinkles} and  Zhang \etal \cite{zhang2020deep} learned high frequency details from normal maps.
In our approach, we cascade multiple geometric features as shape descriptors embedded in geometry images, including spatial information of the displacement, the normal and temporal information of the velocity.

\subsection{\clr{Deep CNN-Based Super-Resolution}}
In the area of single image super-resolution (SISR), deep learning-based techniques  \clr{\cite{ledig2017photo,zhang2018residual,dong2016image, liu2019adaptive, yue2020reference}} have achieved significant breakthroughs in recent years. 
Convolutional layers \cite{dong2016image} are proposed to be more efficient instead of fully-connected structures in SISR, and later on are extended to deep networks using various upsampling layers, \clr{e.g.}, transposed layers \cite{dong2016accelerating} and sub-pixel layers \cite{shi2016real}.
For deep networks, \cll{residual learning \cite{ledig2017photo}, dense connections \cite{haris2018deep}, adaptive residual scheme \cite{liu2019adaptive}, and channel attention module \cite{yue2020reference}} are employed to solve the vanishing gradient problem.
Our method likewise uses a deep network with residual dense architecture \cite{zhang2018residual} for its performance and efficiency.
 
In video super-resolution tasks, how to generate temporal consist results is a vital problem. 
One way is using consecutive frames as inputs \cite{kappeler2016video} or recurrently using previously predicted outputs \cite{chu2020learning}.
More recently, the recurrent mechanism has influenced the field of cloth animations.
\clr{For instance}, Santesteban \etal \cite{santesteban2019learning} used recurrent networks based on gated recurrent units to regress garment wrinkles.
Recurrent modules need to predict results sequentially, while our technique processes images individually and parallelly, even in arbitrary order.
An alternative solution is to synthesize single output with specialized loss terms to constrain the  consistency over time.
Loss functions using nearby frames help to alleviate temporal discontinuities in video \cite{bhattacharjee2017temporal}.    
In fluid generation, Xie \etal \cite{xie2018tempogan} utilized a discriminator loss to preserve temporal coherence. 
For cloth wrinkle synthesis, L\" ahner \etal \cite{laehner2018deepwrinkles} proposed a $L1$ loss between the generated normal map and the ground truth at the previous frame. In our work, the cloth meshes are created by physics-based simulation, \clr{and} thus ground truth motion is available.
A kinematic-based loss constraining the estimated velocity and position enables
our network to generate realistic wrinkles while keeping the predictions coherent from frame to frame.
\begin{figure*}[t]
	\centering 
	\includegraphics[width=\linewidth, trim=30 130 50 100,clip]{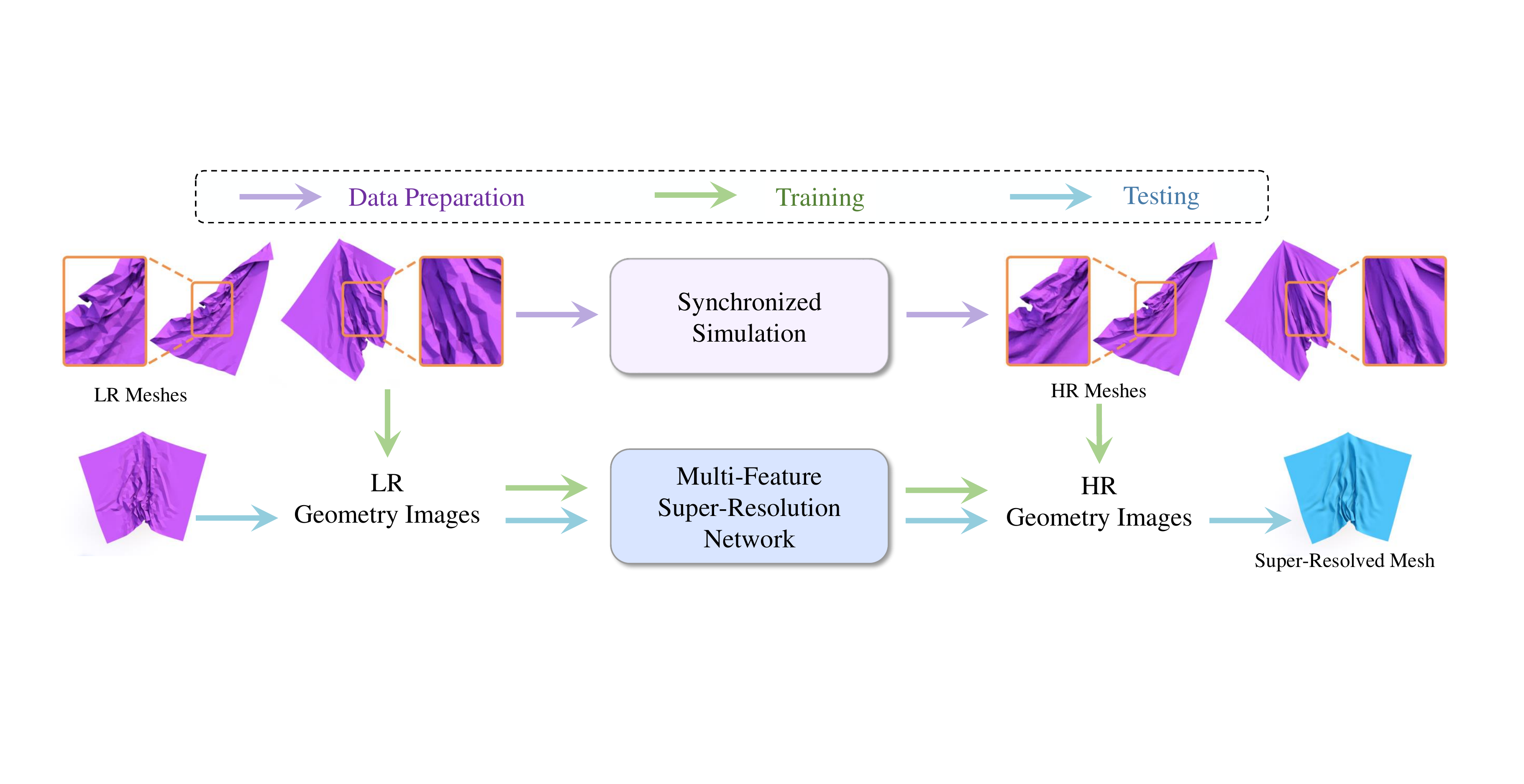} 
	\caption{Pipeline of our Multi-Feature Super-Resolution (MFSR) network for cloth wrinkle synthesis.
		We generate low-resolution (LR) and high-resolution (HR) mesh sequences via synchronized simulation.
		In the training stage, LR and HR meshes are converted into LR and HR geometry images, respectively, encoding multiple features: the displacement, the normal and the velocity of the sampled points.
		Then these features are fed into our MFSR network for training.
		At \clr{the} runtime stage, LR geometry images (converted from the input LR mesh) are super-resolved into HR geometry images,
		which are converted to a detailed mesh.
	}
	\label{fig:system_architecture}
\end{figure*}
\section{Overview}
Our method takes physical simulated LR meshes as input, to infer realistic and consistent HR cloth animations. 
The pipeline of our approach is illustrated in Fig. \ref{fig:system_architecture}.
To generate training data, a pair of LR and HR meshes are simulated synchronously by virtual spring constraints and multi-resolution dynamic models (\clr{Subsection} \ref{Synchronized}).
Thus, the LR and \clr{the} HR meshes are well aligned at the level of large-scale deformation and differ in the wrinkles.
Then the simulated mesh pairs are converted into dual-resolution geometry images (\clr{Subsection} \ref{sec:representation}), with each sample encoding three features: the displacement, the normal and the velocity.
A multi-feature super-resolution network (MFSR) with shared layers and task-specific modules is proposed to super-resolve LR images with details (\clr{Section} \ref{sec:network}).
Based on these features, we design the spatial and temporal loss functions (\clr{Subsection} \ref{sec:loss}) to train our MFSR for detailed and consistent results.
At runtime, the testing LR geometry images (converted from the input LR mesh) are upsampled into HR geometry images, which are then converted to a detailed HR mesh with a refinement step to solve collisions.

\section{\clr{Data Preparation}}

\subsection{\clr{Data Representation and Conversion}}
\label{sec:representation} 

\textbf{Dual-resolution meshes. }
Before executing cloth simulation for data preparation, we need to set the initial rest state of LR and HR meshes.
We obtain the HR mesh by subdividing the edges of the LR one progressively till the desired resolution.
In this work, the number of faces in the HR mesh is 16 times as many as the LR mesh.
With the rest state LR/HR meshes, we create two sets of dual-resolution frame data via physics-based simulation.
The correspondence between them is maintained during the simulation \clr{so that} they exhibit similar large-scale folding behaviors but differ in the fine-level wrinkles.
More details about the synchronized simulation are given in \clr{Subsection} \ref{Synchronized}.

\textbf{Dual-resolution geometry images. }
We convert the paired meshes to dual-resolution geometry images of 9 channels.
The embedded descriptors in our images include \clr{the {displacement} $\bm{d}$, the {normal} $\bm{n}$ and the {velocity} $\bm{v}$.}
Different from the original geometry image paper \cite{gu2002geometry}, we encode the displacements instead of positions, since we are only interested in the intrinsic shape of the mesh, not its absolute spatial locations. 
The displacement is defined as the difference between its position in current frame and that in its starting position.
The vertex normal is computed by the area-weighted average normals of the faces adjacent to this vertex.
Due to the physics-based simulation with \clr{the} fixed time step, the velocity is naturally calculated using the positions between two frames \clr{(the complete calculation of our feature descriptors is provided in Section 1 of the supplementary material)}.
Since these features are not \clr{rotation-invariant}, we calculate a rigid motion transformation \cite{kabsch1978discussion} with rotation $\bm{R}$ and translation $\bm{t}$. Then, we apply ($\bm{R},\bm{t}$)  to displacement, \clr{and $\bm{R}$ is} applied to normal and velocity.
To reduce the computation cost, we only compute the rigid motion of LR meshes and apply the same $\bm{(R,t)}$ to the HR meshes.
To release the internal covariate shift \cite{glorot2010understanding}, these features are normalized into a range of [0, 1].
\begin{figure}[H]
	\centering
	\includegraphics[width=0.98\linewidth]{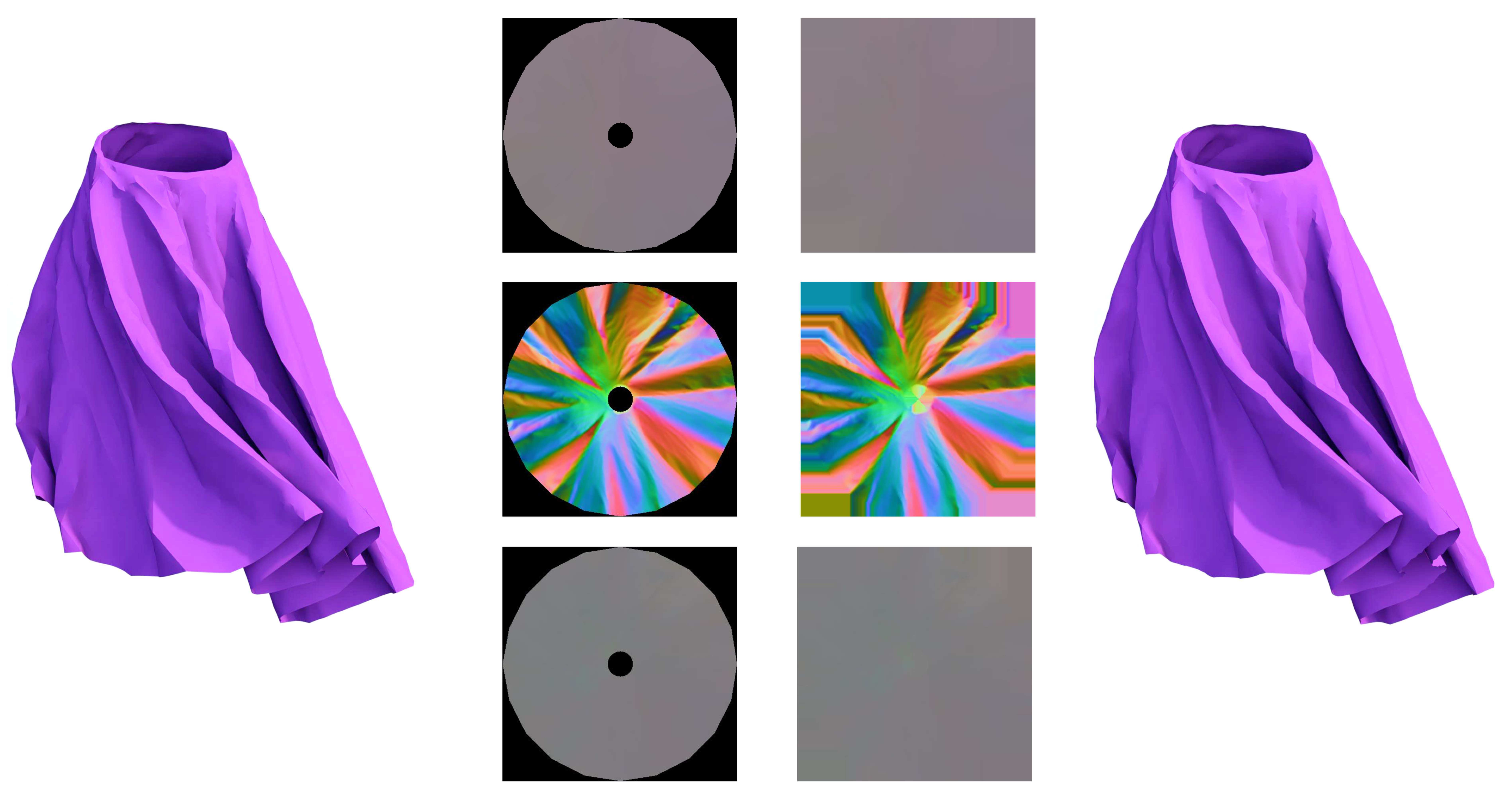}
	\begin{minipage}[b]{0.3\linewidth} 
    \centering (a)
    \end{minipage}
    \begin{minipage}[b]{0.15\linewidth} 
    \centering (b)
    \end{minipage}
    \begin{minipage}[b]{0.15\linewidth} 
    \centering (c)
    \end{minipage}
    \begin{minipage}[b]{0.3\linewidth} 
    \centering (d)
    \end{minipage}
	\caption{Mesh-image conversion. The HR skirt is converted to geometry images with three descriptors: the displacement $\bm{d}$, the normal $\bm{n}$ and the velocity $\bm{v}$.
		For irregular garments, the feature values of sample points outside the mesh but inside the bounding box are zero (black pixels in (b)),
		\clr{and then are padded with the nearest non-zero values (c). Following that, the geometry image of displacement is utilized for mesh reconstruction. (a) HR mesh. (b) Geometry images without padding. (c) Geometry images with padding. (d) Reconstructed mesh.}
	}
	\label{fig:meshimageconversion}
\end{figure}

\textbf{Mesh-to-image conversion. } 
For a mesh in its rest state, we find its bounding box in the 2D material space.
Inside the bounding box, we then sample an array of $m \times n$ points uniformly.
For each sample point inside the mesh, we find the triangle it is located in, and compute its barycentric coordinate (BC) w.r.t. three triangle vertices. 
BC is unchanged even though a triangle deforms during simulation.
When computing features for sample points, BCs are used as weights for interpolating feature values $\bm{(d, n, v)}$ from triangle vertices.
For a mesh whose boundary coincides with the bounding box edge, we do the padding operation along boundaries.
Otherwise, for sample points outside the mesh but inside the bounding box, their feature values are filled with the nearest non-zero pixels similar to replicate padding. 
A long skirt example is given in Fig. \ref{fig:meshimageconversion}.

\textbf{Image-to-mesh conversion. }
After an HR image is synthesized, values in the displacement channels are used to reconstruct the positions of the detailed mesh, while the original topology of that mesh is retained.
Due to the padding operation, every vertex in 2D material space has four nearest non-zero sample points.
We reconstruct the displacements of vertices by bilinear interpolation.
These computed displacements are added to the positions of subdivided mesh vertices in the rest state to obtain wrinkle-enhanced positions.
In the end we apply the inverse of the rigid transformation, computed in the mesh-to-image phase, to new positions.
\clr{As shown in Fig. \ref{fig:meshimageconversion}(d)}, almost no visual differences can be seen.
In our quantitative experiments, the geometric reconstruction error is smaller than 1e-4 meter, measured by the vertex mean square error (VMSE).

\subsection{Synchronized Simulation}
\label{Synchronized}
\begin{figure}[H]
	\centering
\includegraphics[width=\linewidth]{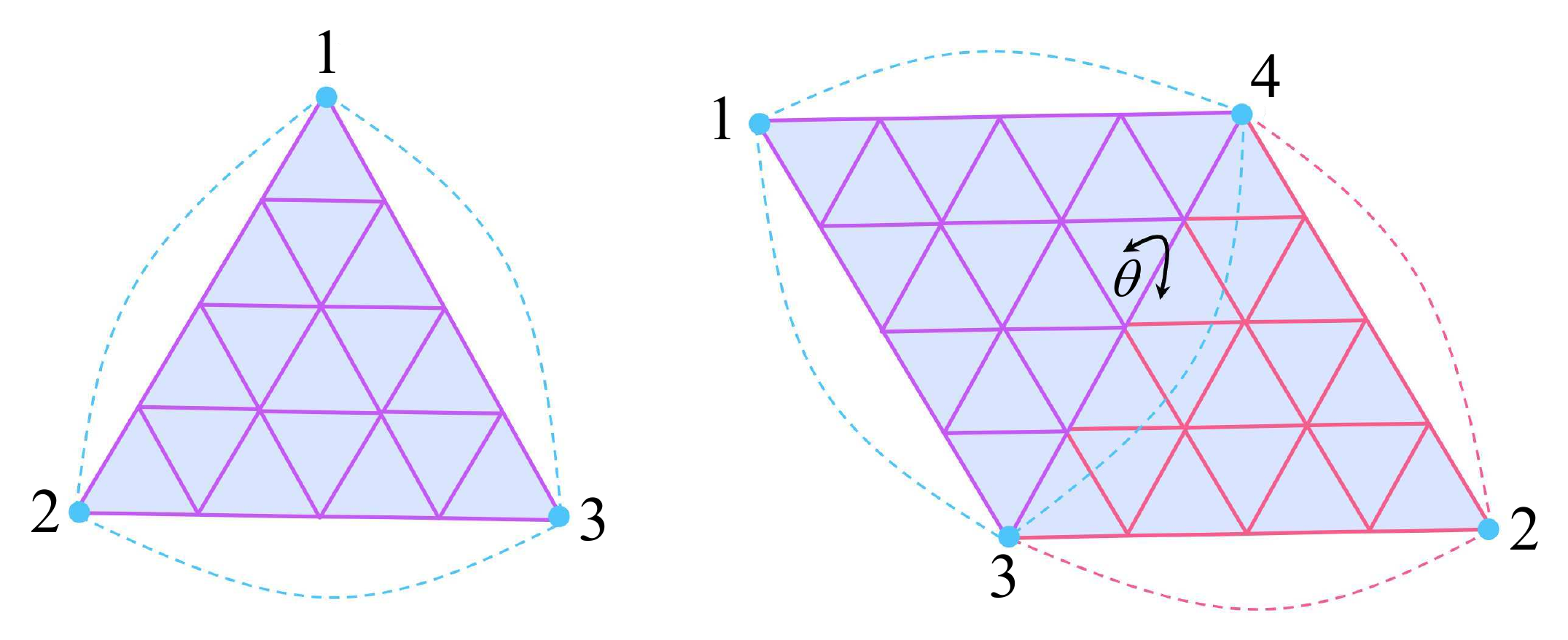}
	\begin{minipage}[b]{0.3\linewidth} 
    \centering (a)
    \end{minipage}
    \begin{minipage}[b]{0.5\linewidth} 
    \centering (b)
    \end{minipage}
	\caption{\small \clr{Multi-resolution dynamic model for tracking. (a) Forces of stretching. (b) Forces of bending.}
	}
	\label{fig:multiresolutiontrack}
\end{figure}

\begin{figure*}[th]
	\centering 
	\includegraphics[width=\linewidth,trim=80 100 140 70,clip]{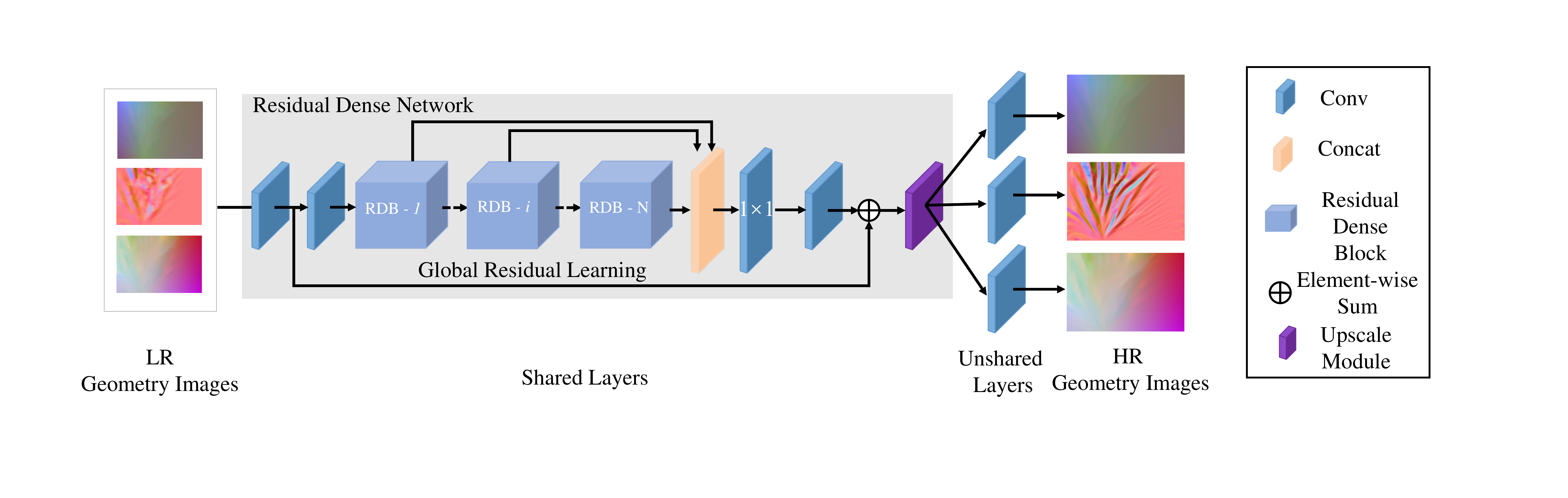}
	\caption{\clr{Architecture of MFSR.} 
		The input and the output are LR/HR images where each pixel is represented as a 9-dimensional feature vector enclosing the displacement, the normal and the velocity in order.
		Conv and Concat refer to convolutional and concatenation layers, respectively. \cll{Here the $1 \times 1$ Conv denotes a special convolutional layer to adaptively fuse the output information of preceding RDBs.}
		The MFSR upscales the LR features with shared and unshared layers to recover HR features with detailed information.
	} 
	\label{fig:multi-tasknetwork}
\end{figure*}

The high-quality training dataset is equally important for data-driven approaches. 
In our case, we need to generate corresponding LR/HR mesh pairs in animation sequences by physics-based simulation.
In image super-resolution tasks \cite{ledig2017photo,dong2016image}, one way to generate training dataset is down-sampling HR images to obtain their corresponding LR ones.
However, down-sampling an HR cloth mesh could cause collisions, even though the HR mesh is collision-free.
Therefore, it is preferred that two meshes are simulated individually, with all collisions being properly resolved.
However, as mentioned in previous works \cite{zurdo2013wrinkles, kavan11physics}, if there is no constraints between two simulations, they will bifurcate to different behaviors because of accumulated higher frequencies generated by finer meshes and numerical errors.
\cl{There are several ways to formulate synchronized constraints such as testing functions \cite{kavan11physics,Bergou2007TRACKS}.
And our implementation enforces virtual spring constraints and uses multi-resolution dynamic models to construct synchronized simulation for HR meshes.
}
 
Our dual-resolution meshes are well aligned in the initial state, because we only add vertices on the edges without changing the mesh shape.
The vertices in an LR mesh, called \clr{feature vertices}, show up in an HR mesh and are used as constraints for synchronized simulation.
We first run coarse cloth simulation and record the positions of all feature vertices at total $N$ frames as ${\bm{p}^{l}_{k}}$, ${k=1, \cdots, N}$, where the superscript $l$ stands for the LR.
While simulating an HR mesh at the frame $k$, virtual springs are added to connect pairs $(\bm{p}^{l}_{k}, \bm{p}^{h}_{k-1})$ of feature vertices between \clr{the} LR mesh at the frame $k$ and \clr{the} HR mesh at the frame $k-1$.
To pull $\bm{p}^{h}_{k-1}$ towards $\bm{p}^{l}_{k}$, we define an internal force following Hooke's law as
\begin{equation}
	\bm{f}_{spring} = - c ( \bm{p}^{l}_{k} - \bm{p}^{h}_{k-1} )\clr{,} \label{con:springforce}
\end{equation}
where $c$ is a spring stiffness constant that can be adjusted depending on how tight the tracking is desired by the user.
A large $c$ results in tight tracking of the feature vertices, but not for other vertices.
As a side effect the simulated HR mesh has many annoying ``spikes".

Thus, we add a multi-resolution dynamic model to cooperate with virtual springs.
Given an HR mesh at level $H_0$ (shown as the solid lines in Fig. \ref{fig:multiresolutiontrack}), we construct an LR triangle mesh at level $H_1$ (the dashed triangle in Fig. \ref{fig:multiresolutiontrack}).
The mesh in $H_1$ connects the feature vertices by retaining the topology of the LR mesh.
In finite-element simulations, the constitutive model includes internal cloth forces supporting behaviors such as anisotropic stretch or compression \cite{muller2004interactive} and surface bends \cite{bridson03wrinkles} with damping \cite{Narain2012AAR}. 
For a triangle in the coarse mesh at level $H_1$, the in-plane stretching forces $\bm{f}^{s}_{1} = (\bm{f}^{s}_{11}, \bm{f}^{s}_{12}, \bm{f}^{s}_{13})$ at three vertices are measured by a corotational finite-element approach \cite{muller2004interactive}.
\clr{The} bending forces for two adjacent triangles are added using a discrete hinge
model based on dihedral angles, denoted as $\bm{f}^b_1 = (\bm{f}^b_{11}, \bm{f}^b_{12}, \bm{f}^b_{13}, \bm{f}^b_{14})$.
The triangles in the fine level $H_0$ have the same force patterns $\bm{f}^{s}_{0}$ and $\bm{f}^b_0$ imposed on all particles (including feature vertices).
All stretching and bending forces are added accompanying damping forces.
In addition, our two-level dynamic models are independent of the force implementations, and would also work with other triangular finite-element methods.
As a result, the feature vertices in multi-resolution dynamic models receive the stretch forces from both $\bm{f}^{s}_{0}$ and $\bm{f}^{s}_{1}$, \clr{and the bending forces from both $\bm{f}^{b}_{0}$ and $\bm{f}^{b}_{1}$.} 
The rest vertices are only imposed on the forces at level $H_1$.
With the two-hierarchy dynamics model, modest virtual spring coefficients can make the HR mesh keep pace with the LR mesh in simulation.

\section{Multi-Feature Super-Resolution Network}
\label{sec:network}
In this section, we introduce our MFSR architecture based on the RDN, as well as the loss functions taking spatial and temporal features into account to improve wrinkle synthesis capability.

\subsection{MFSR Architecture}
We now introduce our MFSR architecture for the image SR tasks of multiple features.
With LR/HR images of the form $(\bm{d}, \bm{n},  \bm{v})^{l}$ and $(\bm{d}, \bm{n}, \bm{v})^{h}$, our MFSR learns the mappings of different features by image SR networks.
One standard methodology is single task learning, which means learning one task at a time.
However it ignores a potentially rich source of information available in other tasks.
Another option is multi-task learning, which achieves inductive transfer between tasks, with the goal to leverage additional sources to improve the performance of the target task \cite{caruana1997multitask}.
Our MFSR is a multi-task architecture, \clr{consisting} of two components: a single shared network, and three task-specific networks. 
The shared network is designed based on the SR task, whilst
each task-specific network consists of a set of convolutional modules, which link with the shared network.
Therefore, the features in the shared network, and the task-specific networks, can be learned jointly to maximise the generalisation of the shared representation across multiple SR tasks, simultaneously maximising the task-specific performance.

\begin{figure}[H]
	\centering
	\includegraphics[width=1.0\linewidth]{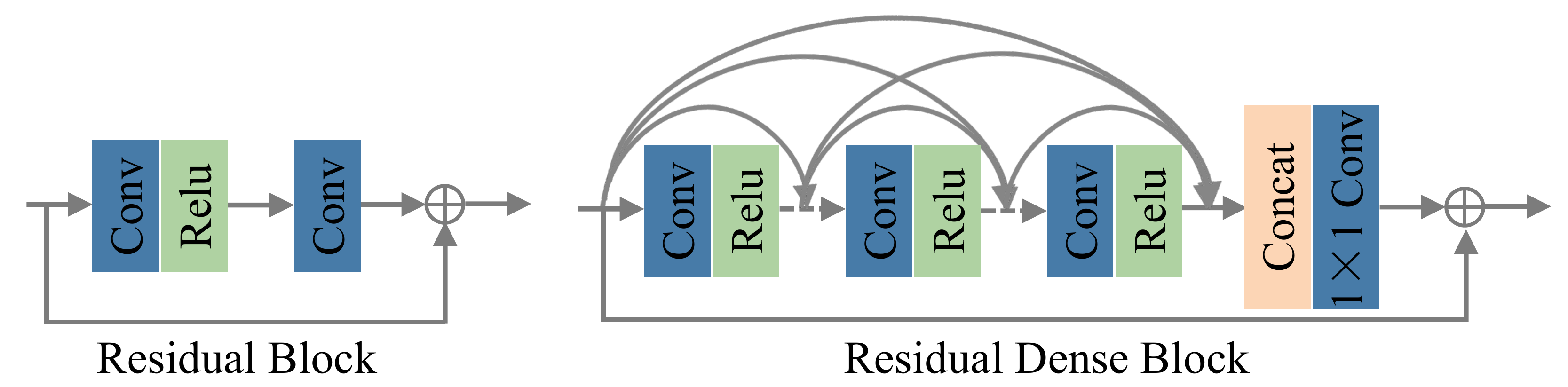}
	\begin{minipage}[b]{0.3\linewidth} 
    \centering (a)
    \end{minipage}
    \begin{minipage}[b]{0.6\linewidth} 
    \centering (b)
    \end{minipage}
	\caption{\small \clr{Two network structures used in our MFSR.
		(a) Residual block in \cite{ledig2017photo}.
		(b) Residual dense block in \cite{zhang2018residual}.
	}}
	\label{fig:RDB}
\end{figure}

Fig. \ref{fig:multi-tasknetwork} shows a detailed visualisation of our MFSR based on residual dense blocks (RDB) \cite{zhang2018residual}.
In the shared network, the image SR model consists of four parts: \clr{shallow feature extraction, basic blocks, dense feature fusion, and finally upsampling}.
We use two convolutional layers to extract shallow features, followed by the RDB \cite{zhang2018residual} as the basic blocks, then dense feature fusion to extract hierarchical features, and lastly one bilinear upsampling layer to upscale the height and width of the LR feature maps by 4 times.
Different from general SR tasks, we find that pixel shuffle and deconvolution methods cause apparent checkboard artifacts, \clr{thereby we use the} bilinear method.
For basic blocks in our SR network, we employ RDB instead of residual blocks used in \cite{chen2018synthesizing}.
\clr{As shown in Fig. \ref{fig:RDB}(a)}, a residual block learns a mapping function with reference to its input, \clr{and} therefore can be used to build deep networks to address the problem of vanishing gradients.
However, in the residual block a convolutional layer only has direct connection to its precedent layer, neglecting to make full use of all preceding layers.
To exploit all the hierarchical features, we choose RDB \clr{(see in Fig. \ref{fig:RDB}(b)) that consists} of densely connected layers for global feature combination and local feature fusion with local residual learning.
More details about RDB are given in \cite{zhang2018residual}.
In each task-specific network, we utilize one convolutional layer to map the extracted local and global features to each upsampled descriptor $\bm{d}^{s}, \bm{n}^{s}$,  and $\bm{v}^{s}$, respectively.
\subsection{Spatial and Temporal Losses}
\label{sec:loss}
In order to learn the spatial details and temporal consistency of the underlying HR meshes, our MFSR is trained by minimizing the following loss functions for mesh features.
A baseline mean square error (MSE) reconstruction loss is defined as
\begin{equation}
	\mathcal{L}_{d} = ||\bm{d}^{h} - \bm{d}^{s}||^2 ,
\end{equation}
\cl{where $\bm{d}^{h}$ and $\bm{d}^{s}$ stand for the ground truth HR and the synthesized SR displacement images, respectively.}
This displacement loss term is able to obtain a smooth HR result with given low frequency information.

To extend the loss into wrinkle feature space, a novel \clr{${L}_{2}$} loss for normal is introduced:
\begin{equation}
	\mathcal{L}_{n} = ||\bm{n}^{h} - \bm{n}^{s}||^2\clr{,}
\end{equation}
\cl{where $\bm{n}^{h}$ denotes the ground truth normal image and $\bm{n}^{s}$ denotes the superresolved normal image.}
The normal feature is directly related to the bending behavior of cloth meshes.
This loss term encourages our model to learn the fine-level wrinkle features so that the outputs can stay as close to the ground truth as possible.
In our experiments it aids the networks in creating realistic details.

The above two loss terms are utilized to reconstruct high-frequency details exclusively from spatial statistics.
To improve the consistency for animation sequences, we should also take the temporal coherence into account.
The vertex velocities of every animation frame contribute a velocity loss of the form
\begin{equation}
	\mathcal{L}_{v} = ||\bm{v}^{h} - \bm{v}^{s}||^2,
\end{equation}
\cl{where $\bm{v}^{h}$ denotes the ground truth velocity image and $\bm{v}^{s}$ denotes the synthesized velocity image.}

In addition, we minimize a kinematics-based loss in the training stage, to constrain the relationship between synthesized velocities and displacements (please refer to the supplementary material for the detail derivation) as
\begin{equation}
	\mathcal{L}_{kine} = \sum_{k=1}^{n}||\bm{R}^{-1}(\bm{d}^{s}_{k}  -(\bm{d}^{s} + (\sum_{j=1}^{k} \bm{v}^{s}_{k-j} )*\Delta t) ||^2 , \label{con:kineloss}
\end{equation}
where $n$ is the length of frames associated to the input frame, and $\Delta t$ represents the time step between consecutive frames. 
\cl{
$\bm{R}$ is the precomputed rotation part in the rigid motion transformation for input frame.}
This kinematics-inspired loss term can improve the consistency between the generated cloth animations.

The overall loss of our MFSR is defined as
\begin{equation}
	\mathcal{L}_{all} = w_{d}\mathcal{L}_{d} +  w_{n}\mathcal{L}_{n} + w_{v}\mathcal{L}_{v} + w_{kine}\mathcal{L}_{kine}\clr{,} \label{con:lossall}
\end{equation}
which is a linear combination of spatial smoothness, detail similarity, temporal consistency and kinematic loss terms with the weight factors $w_{d}, w_{n}, w_{v}$, and $w_{kine}$.
As for back propagation, each loss term propagates backwards through the task-specific layer independently.
In the shared layers, parameters are updated according to the total loss $\mathcal{L}_{all}$.
As a result, the gradient of loss functions from multiple SR tasks will pass through the shared layers directly,
and learn a common representation for all related tasks. 

The reconstructed meshes converted from super-resolved images by the above network may suffer from penetrations with obstacles or self-collision. 
For interactions with human or balls, we adopt a fast refinement method \cite{wang2019learning} to push the cloth vertices colliding with the obstacles outside \clr{and} meanwhile preserving the local wrinkle details.
As for self-collision, 
\cl{since the runtime simulation of an LR mesh is collision-free, we interpolate the vertices in the penetrating area between LR meshes and the reconstructed ones to guarantee to resolve all collisions.
In implementation, we use the bisection method \cite{burden2011numerical} to search for a close-to-optimal interpolation weight. 
We do the bisection several times and then take the last collision-free state as the interpolation result (see Fig. \ref{fig:collisonhandling}). 
It is not necessary to let all vertices of the whole mesh get involved in the position interpolation.
Instead, only the vertices involved in the intersections are of our interests. 
These vertices can be specified by a discrete collision detection process and grouped into impact zones as done in \cite{ye17unified}. 
Position interpolations are performed per zone, and each zone has different interpolation weights. 
In this way, the synthesized meshes are least affected by the collision handling. 
 \begin{figure}[H]
	\centering  
	\includegraphics[width=1\linewidth, trim=100 10 100 10, clip]{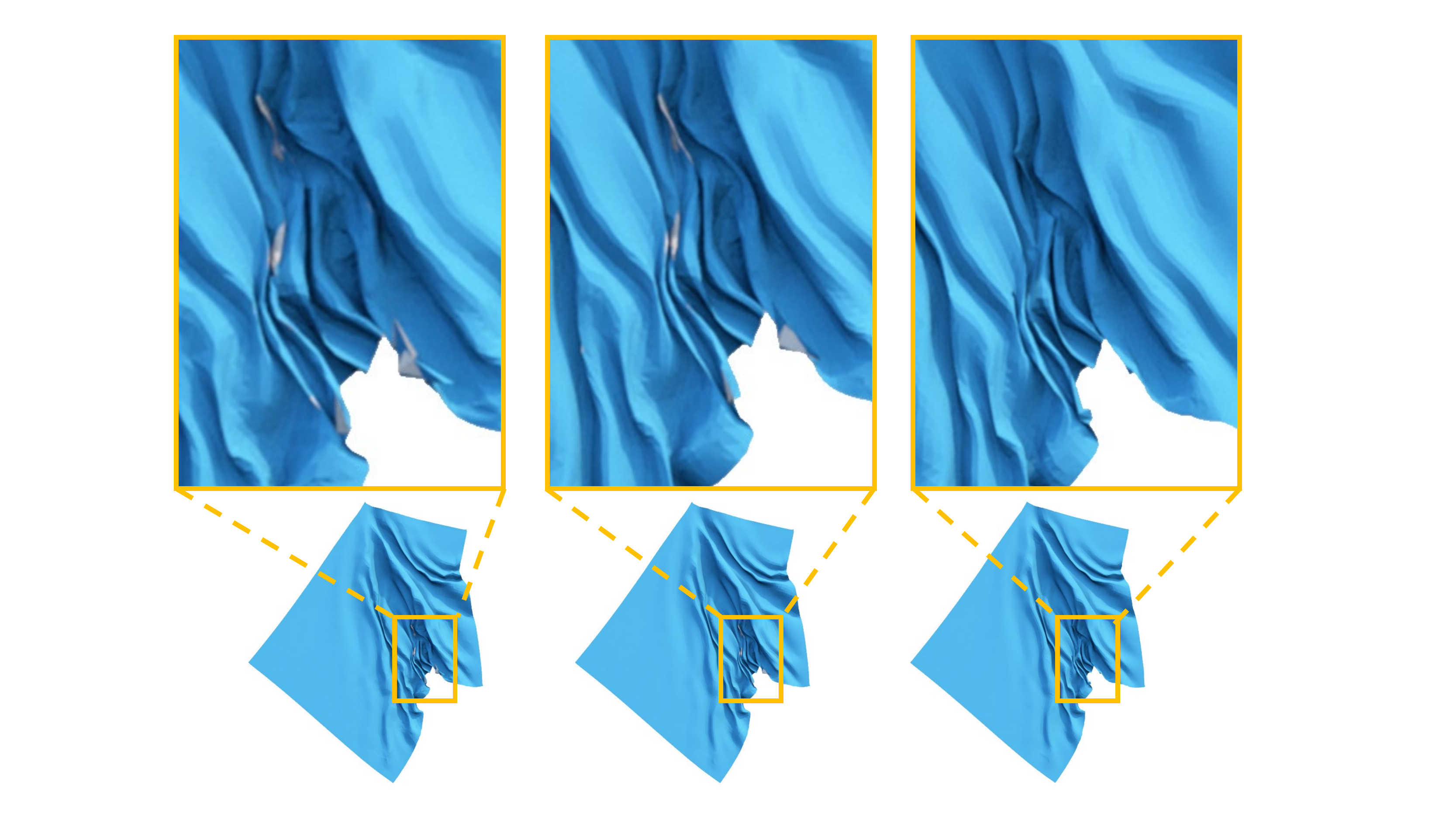}
	\begin{minipage}[b]{0.3\linewidth} 
    \centering (a)
    \end{minipage}
    \begin{minipage}[b]{0.3\linewidth} 
    \centering (b)
    \end{minipage}
    \begin{minipage}[b]{0.3\linewidth} 
    \centering (c)
    \end{minipage}
	\caption{\small \clr{Given a super-resolved cloth with self-collision, the collision solving method is utilized to untangle the intersection regions after two steps. (a) Super-resolved cloth with collision. (b) Collision solving cloth after one step. (c) Collision-free cloth after two steps.} 
	}
	\label{fig:collisonhandling}
\end{figure} 
}

\section{Implementation}
We describe the details of the data generation and the network architecture in this section.

\subsection{\clr{Data Generation}}
\label{sec:dataset}
\begin{figure}[H]
	\centering 
\includegraphics[width=\linewidth]{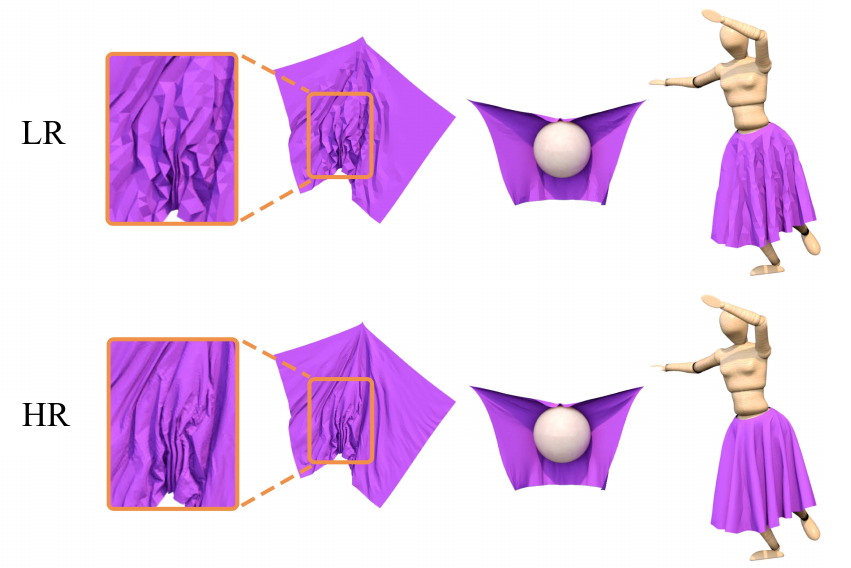}
    \begin{minipage}[b]{0.6\linewidth} 
    \centering (a) 
    \end{minipage}
    \begin{minipage}[b]{0.2\linewidth} 
    \flushleft (b) 
    \end{minipage}
    \begin{minipage}[b]{0.15\linewidth} 
    \centering (c) 
    \end{minipage}
	\caption{\small Three constructed datasets for evaluation. The top and bottom rows show the LR and HR cloth meshes, respectively. (a) DRAPING. (b) HITTING. (c) SKIRT.
	}
	\label{fig:tablecloth_dataset}
\end{figure}

We construct three datasets using a tablecloth model and a skirt model with character motions.
The two models are regular and irregular garment shapes, respectively.
The meshes in each dataset are simulated from a fixed template model.
For the tablecloths, we generate two datasets, called DRAPING and HITTING (see Fig. \ref{fig:tablecloth_dataset}). 
The DRAPING dataset is created by randomly handling one of the topmost vertices of the tablecloth and letting the fabric fall freely.
It contains 13 simulation sequences, each with 400 frames. 10 sequences are randomly selected for training and remaining 3 sequences are for testing.
In addition to simulating a piece of tablecloth in a free environment, we also construct a HITTING dataset where a sphere interacts with the tablecloth.
Specifically, we select spheres of different sizes to hit the tablecloth back and forth at different locations, and obtain a total of 35 simulation sequences, with 1,000 frames for each sequence.
We randomly select 27 sequences for training and 8 sequences for testing.
The SKIRT dataset is created by the long skirt garments worn by an animated character (shown in Fig. \ref{fig:tablecloth_dataset}).
A mannequins has rigid parts as \cite{Narain2012AAR} and is driven by publicly available motion capture data from CMU \clr{\footnote{http://mocap.cs.cmu.edu}}.
We select dancing motions including 7 sequences (in total 30,000 frames), in which 5 sequences are randomly selected for training and 2 sequences are for testing.
To simulate cloth stably, we interpolate 8 times between two adjacent motions from the original data. 

We apply the ARCSim engine \cite{Narain2012AAR} to produce all simulations with remeshing disabled.
A material called the Gray Interlock is adopted for its anisotropic behaviors, from a library of measured cloth materials \cite{Wang2011DEM}.
To meet a collision-free initial state for skirts, we first manually put the skirt on a template mannequin (T pose), \clr{and} then interpolate 80 frames between the T pose and the initial poses of motion sequences.
In addition, for synchronized simulation, we set the spring stiffness constant $c = 10$ in the equation (\ref{con:springforce}).

\cll{
\begin{table*}[]
	\caption{ Statistics and Timing (sec/frm) of the Tablecloth and Skirt Testing Examples}
	\label{table:runtime}
	\centering
	\begin{tabular}{cccccccccc}
		\hline
		 \multirow{3}{*}{Dataset} &  &  &   & \multirow{3}{*}{Ours}& \multirow{3}{*}{\clr{Speedup}}&\multicolumn{4}{c}{\clr{Our Components}}              \\ \cline{7-10} 
		  &  \#verts  & \#verts  & \clr{Tracked} &   &     & \clr{Coarse} & \clr{Mesh/Image} & \clr{Synthesizing} & \clr{Refinement} \\
		&   LR      & HR      & \clr{Sim.}     &       &         & Sim.   & Conversion & (GPU)        &            \\  \hline 
		DRAPING     & 749     & 11,393   & 4.27    & 0.345 & \textbf{12}      & 0.129  & 0.089      & 0.0553       & 0.0718     \\
		HITTING   & 749     & 11,393   & 4.38    & 0.341 & \textbf{13}      & 0.135  & 0.109      & 0.0531       & 0.0434     \\
		SKIRT     & 1,303    & 19,798   & 10.23   & 0.709 & \textbf{14}     & 0.227  & 0.18       & 0.0281       & 0.274      \\  \hline 
	\end{tabular}
	\begin{tablenotes}
		\item \footnotesize Note: ``\#'' means ``number of'', ``Sim.'' is abbreviated for ``simulation''.
	\end{tablenotes}
\end{table*} 
}

\subsection{\clr{Network Architecture}} 
\label{sec:networkparameter}
For different datasets, we train each model separately. 
Our proposed MFSR consists of shared and task-specific layers.
The shared network has 16 identical RDB \cite{zhang2018residual}, where six of them are densely connected layers for each RDB, and the growth rate is set to 32. 
The basic network settings, such as the convolutional kernel and \clr{the} activation function, are set according to \cite{zhang2018residual}.
For the upscaling operation, \clr{i.e.}, from the coarse resolution features to fine ones, we consider several different mechanisms, \clr{e.g.}, pixel shuffle module \cite{shi2016real}, deconvolution, nearest and bilinear, and finally choose the bilinear upscaling layer because it can prevent checkerboard artifacts in the generated meshes. 
In our upsampling network, the upscale factor is set to 4.
The upscale factor (in one dimension) for corresponding meshes is set to be as close to 4 as possible.
For example, the LR and the HR tablecloth meshes have 749 and 11,393 vertices, respectively, \clr{and the latter is} roughly 16 times as many as the former.
Converting meshes to images, we set the size of LR images in tablecloth to be $192 \times 128$, and HR ones $768 \times 512$.
The image aspect ratio is the same to the \clr{UV} proportion in material space to achieve uniform sampling.  

We implement our network using PyTorch 1.0.0.
In each training batch, we randomly extract 16 LR/HR pairs with the size of $72 \times 72$ and $288 \times 288$ as input.
Adam optimizer \cite{kingma2014adam} is used to train our network, and its $\beta_1$ and $\beta_2$ are both set to 0.9.
The base learning rate is initialized to 1e-4, and is divided by 10 every 20 epochs. To avoid \cll{the} learning rate becoming too small, we fix it after 60 epochs. The training procedure stops after 120 epochs \cl{and takes about a day and a half}.
In all our experiments, we set the length of the input frames $n = 3$ for the kinematics-based loss in the equation ($\ref{con:lossall}$).
Besides, we set the weights $w_{d} = 0.9, w_{n} = 0.03, w_{v} = 0.03$ and $w_{kine} = 0.03$ in the equation (\ref{con:lossall}).

\section{Results and Evaluations}
In this section, we evaluate the results obtained with our method both quantitatively and qualitatively.
The runtime performance and visual fidelity are demonstrated with various scenes: draping and hitting tablecloths, and long skirts worn by animated character, separately.
We compare our results against simulation methods and demonstrate the benefits of our method for cloth wrinkle synthesis.
The effectiveness of our network components is also analyzed, for various loss functions and network architectures. 

\subsection{Runtime Performance}
We implement our method on a 2.50GHz Core 4 Intel CPU for coarse simulation and mesh-image conversion,
and a NVIDIA GeForce\textsuperscript{\textregistered}~GTX 1080Ti GPU for image synthesizing.
Table~\ref{table:runtime} shows average per-frame execution time of our method for \clr{different garment} resolutions.
The execution time contains four parts: coarse simulation, mesh/image conversion, image synthesizing, and refinement. 
For reference, \clr{we also calculate the simulation timings} of a CPU-based implementation of tracked high-resolution simulation using ARCSim \cite{Narain2012AAR}.
Our algorithm is averagely 13 times faster than the tracked simulation.
The low computational cost of our method makes it suitable for the interactive applications. 

\subsection{Wrinkle Synthesis Results and Comparisons}

\begin{figure}[H]
	\centering
    \includegraphics[width=\linewidth]{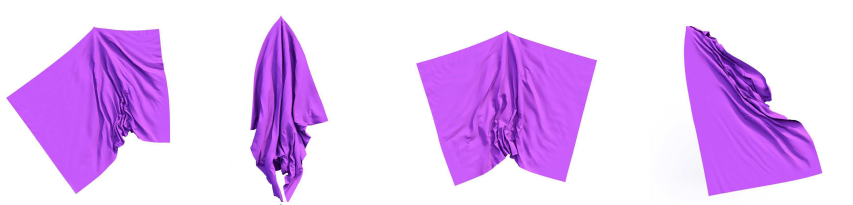} \\
    \centering (a) \\
    \includegraphics[width=\linewidth]{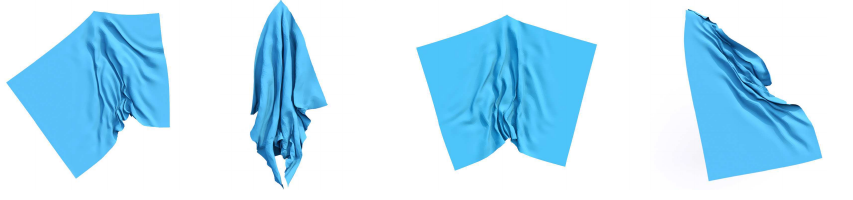} \\
    \centering (b) \\
	\caption{\small \cl{The super-resolved results of our method are able to produce as many wrinkles as the ground-truth tracked HR simulation. (a) Ground Truth. (b) Ours.}
	}
	\label{fig:hangdatasetcompare}
\end{figure}

\textbf{Generalization to new hanging.}
We use the training data in the DRAPING dataset to learn a synthesizer, \clr{and} then evaluate the generalization to new hanging vertices.
Fig. \ref{fig:hangdatasetcompare} shows the deformations of tablecloths of three test sequences in the DRAPING dataset.
We compare our results with the HR meshes of tracked physics-based simulation.
Our approach successfully produces the realistic and abundant wrinkles in different deformation sequences.
\clr{For instance, tablecloths appear middle and small wrinkles when falling from different directions.}
\begin{figure}[H]
	\centering
\includegraphics[width=\linewidth]{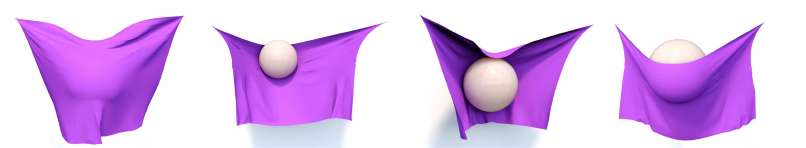} \\
    \centering (a) \\
    \includegraphics[width=\linewidth]{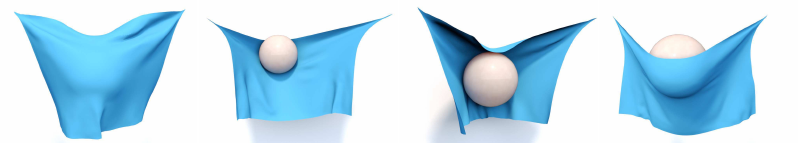} \\
    \centering (b) \\
	\caption{\small Comparison between the ground-truth tracked simulation and our super-resolved meshes, on testing animation sequences in \clr{the} HITTING dataset.
		Our method succeeds to predict the small and mid-scale wrinkles of the garments with 12 times faster running speed than physic-based ones. \clr{(a) Ground Truth. (b) Ours.}
	}
	\label{fig:crashdatasetcompare}
\end{figure}
\textbf{Generalization to new balls.}
Fig. \ref{fig:crashdatasetcompare} presents the performance of our algorithm in the HITTING dataset, which illustrates the performance when generalizing to new crashing balls of various sizes and initial positions.
We show four test examples comparing the ground-truth HR of the tracked simulation with our method.
For testing, the initial positions of balls are set to four different places which are unseen in training data.  
Additionally, {in the last two columns of Fig. \ref{fig:crashdatasetcompare}}, the diameter of the ball is set to 0.5 which is a new size not used for training.
When various sizes of balls crash into the cloth in different positions, our method can successfully predict the plausible wrinkles, with 12 times faster running speed than physics-based simulation.

\begin{figure}[H]
	\centering 
\includegraphics[width=\linewidth]{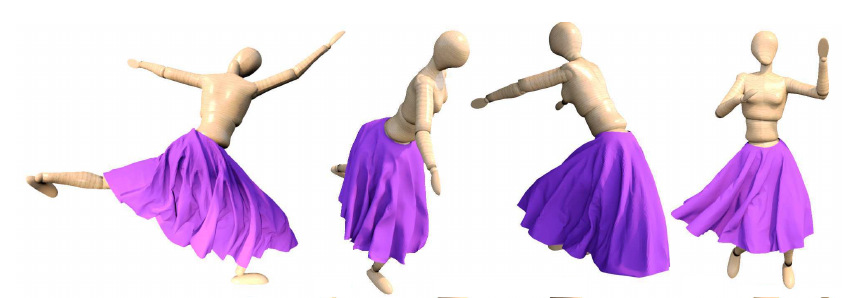} \\
    \centering (a) \\
    \includegraphics[width=\linewidth]{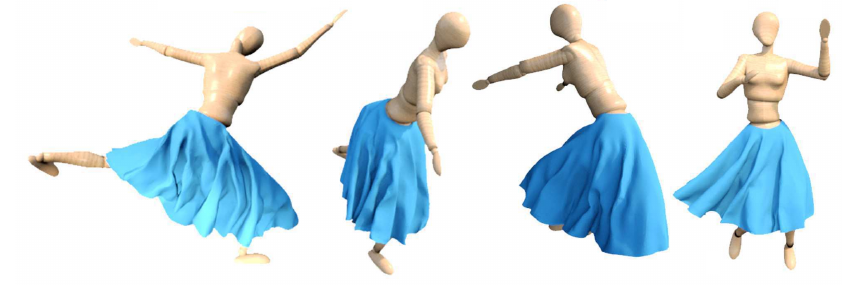} \\
    \centering (b) \\
	\caption{\small Comparison between ground-truth tracked simulation and our super-resolved meshes, on the test frames in the SKIRT dataset. \cl{Our method succeeds to predict the dynamic wrinkles of the long skirts as realistic as physic-based ones.} \clr{(a) Ground Truth. (b) Ours.}
	}
	\label{fig:skirtdatasetcompare}
\end{figure}

\textbf{Generalization to new motions.}
In Fig. \ref{fig:skirtdatasetcompare}, we show the deformed long skirt produced by our approach on the mannequins while changing various poses over time.
The human poses are from two testing motion sequences $05\_04$ in the subject of modern dance and $55\_02$ in the subject of lambada dance \clr{\footnote{http://mocap.cs.cmu.edu}}.
We visually compare the results of our algorithm with the ground-truth simulation.
The mid-scale wrinkles are successfully predicted by our approach when generalizing to various dancing motions not in the training set.
For instance, in the first column of Fig. \ref{fig:skirtdatasetcompare}, the skirt slides forward and forms plausible wrinkles due to an extended and straight leg caused by the character pose of sideways arabesque.
As for dancing sequences, please see the accompanying video for more animated results and further comparisons. 

\textbf{Comparison with other methods.}
Given detailed meshes simulated by the physics-based technique as ground truth, we compare our results with our implementation of a CNN-based method \cite{chen2018synthesizing} and a conventional machine learning-based method \cite{zurdo2013wrinkles}.
The performance is evaluated on the Tablecloth dataset combining DRAPING and HITTING by a single network.
The partition of the dataset for training and testing and the training parameters of our method are the same with the setting illustrated in \clr{Subsection} \ref{sec:networkparameter}.
\begin{figure*}[tb]
\includegraphics[width=1.000000\linewidth]{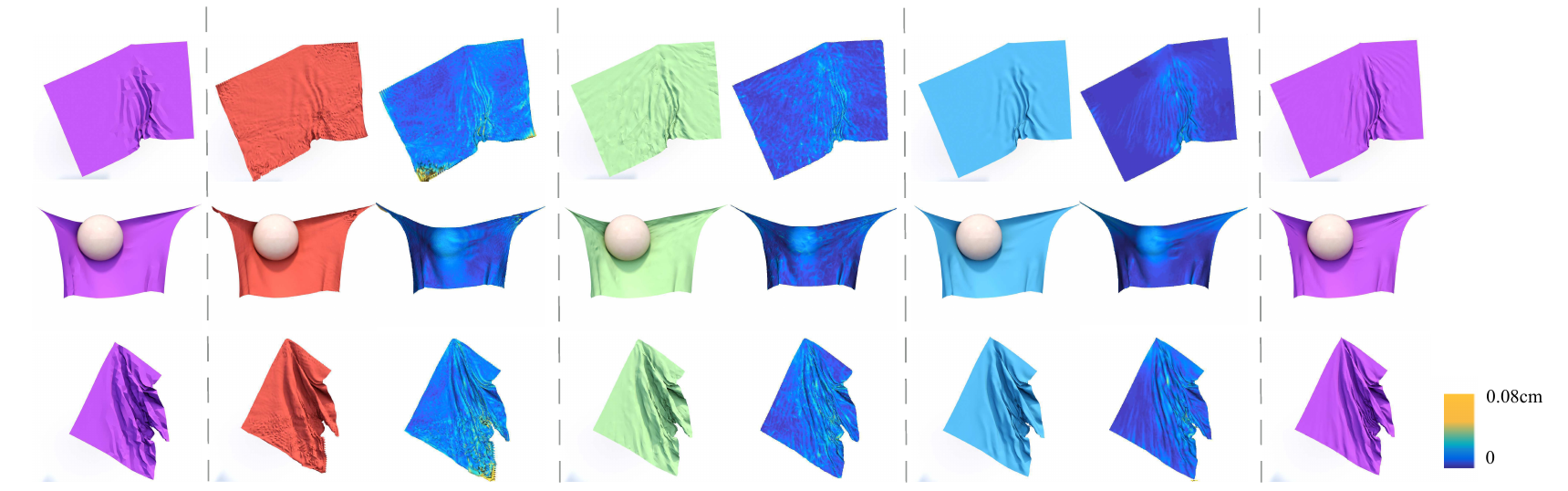} \\
\begin{minipage}[b]{0.11\linewidth} 
\centering (a) 
\end{minipage}
\begin{minipage}[b]{0.22\linewidth} 
\centering (b)  
\end{minipage} 
\begin{minipage}[b]{0.22\linewidth} 
\centering (c)  
\end{minipage} 
\begin{minipage}[b]{0.22\linewidth} 
\centering (d) 
\end{minipage} 
\begin{minipage}[b]{0.15\linewidth} 
\centering (e) 
\end{minipage}
	\caption{\small Comparison of the reconstruction results for unseen data in the Tablecloth dataset combining DRAPING and HITTING.
		The reconstruction accuracy is qualitatively showed as a difference map. 
		Reconstruction errors are color-coded and warmer colors indicate larger errors. Our method leads to significantly lower reconstruction errors. 
		\clr{
		(a) Input coarse meshes.
		(b) Results of Chen \etal \cite{chen2018synthesizing}.
		(c) Results of Zurdo \etal \cite{zurdo2013wrinkles}.
		(d) Our results.
		(e) Ground truth.}
	}
	\label{fig:comparewithcasa}
\end{figure*} 

We train the network of Chen \etal \cite{chen2018synthesizing} with the same setting reported in their paper.
The peak signal-to-noise ratio (PSNR) and \clr{the} vertex-wise mean square error (VMSE) are used to evaluate the quality of reconstructions, quantitatively.
As shown in Table~\ref{table:compare_casa}, our MFSR gains better performance than \cite{chen2018synthesizing} with a higher PSNR and a lower VMSE.
\cll{We also show comparison results in Fig. \ref{fig:comparewithcasa}, with color coding to highlight the distance between the predicted results and the ground truth.
Feeding the coarse meshes, our MFSR successfully produces rich and consistent wrinkles thanks to displacement, normal and velocity features with corresponding loss functions. 
\clr{Chen \etal \cite{chen2018synthesizing} are} able to predict the overall wrinkles of the draping and \clr{hitting tablecloth. However, there} are unsmooth triangles leading to spatial and temporal artifacts since no temporal loss modules \clr{are applied}. This leads to unstable animations, please refer to the accompanying video. 
The difference map also clearly indicates that the results of \clr{Chen \etal \cite{chen2018synthesizing} highlight} the bottom left and right corners and wrinkle lines, where our method leads to significantly lower reconstruction errors.
Our MFSR model provides better visual quality for wrinkle synthesis and the generated HR results look closer to the ground truth.}
\cll{We further make comparison with state-of-the-art conventional machine learning-based method \cite{zurdo2013wrinkles} (not deep learning-based) for cloth wrinkle synthesis. 
Since images are not utilized by Zurdo \etal \cite{zurdo2013wrinkles}, we use a metric, vertex-wise mean square error (VMSE), for quantitative comparison.
As shown in Table~\ref{table:compare_casa}, \clr{our results gain lower VMSE than the results of Zurdo \etal \cite{zurdo2013wrinkles}.}
The results of \cite{zurdo2013wrinkles} show inaccurate and unexpected rough wrinkles different from the ground truth, \clr{e.g.}, the left and right sides in the first row of Fig. \ref{fig:comparewithcasa} (c).
Their conventional example-based algorithm is reliant on a very limited number of examples and parameters.
Thus their method can reconstruct good results while the testing samples are near the example poses, but may cause unexpected artifacts for more diverse testing data.
The accompanying video show that the results of \cite{zurdo2013wrinkles} have small artifacts and are lack of temporal coherence.
This is due to their designed mesh descriptor, the edge ratio between current and the rest state, which only encodes quasistatic wrinkle formation without dynamic features.
Our method does not have such drawbacks with the help of spatial and dynamic descriptors as inputs and temporal loss constraints.
Notice how our method successfully \clr{reconstructs} the stable and realistic cloth animations.
Besides, collisions are not handled in their work. 
Our approach solves cloth penetrations with controllable cost (see in Table~\ref{table:runtime}).}

\begin{table}[H]
	\centering
	\caption{Comparison of Pixel-wise and Vertex-wise Error Values (PSNR/VMSE) on the Tablecloth dataset (DRAPING and HITTING Jointly)}.
	\begin{tabular}{cccc}
		\hline
		\multirow{2}{*}{Dataset} & \multirow{2}{*}{Methods} & \multicolumn{2}{c}{Metrics} \\ \cline{3-4}
		&                          & PSNR $\uparrow$  & VMSE $\downarrow$              \\ \hline 
		\multirow{3}{*}{DRAPING}   & Chen \etal \cite{chen2018synthesizing}  &  59.07 & 4.19e-4        \\   
		& Zurdo \etal \cite{zurdo2013wrinkles} & - & 1.30e-4                     \\
		& Ours                         & \textbf{68.91} & \textbf{7.09e-5} \\ \hline
		\multirow{3}{*}{HITTING}   & Chen \etal \cite{chen2018synthesizing}  &  59.15 & 1.17e-4        \\   
		& Zurdo \etal \cite{zurdo2013wrinkles} & - & 5.78e-5                    \\
		& Ours                         & \textbf{72.25} & \textbf{4.69e-5} \\ \hline
	\end{tabular}
	\begin{tablenotes}
		\item \footnotesize Note: ``$-$'' means a not evaluated PSNR value since images are not utilized by Zurdo \etal \cite{zurdo2013wrinkles}.
		The number in bold indicates the best performance. ``$\uparrow$'' means higher is better. ``$\downarrow$'' means lower is better.
	\end{tablenotes}
\label{table:compare_casa}
\end{table}

\cll{
\begin{figure}[H]
	\centering
	\includegraphics[width=\linewidth]{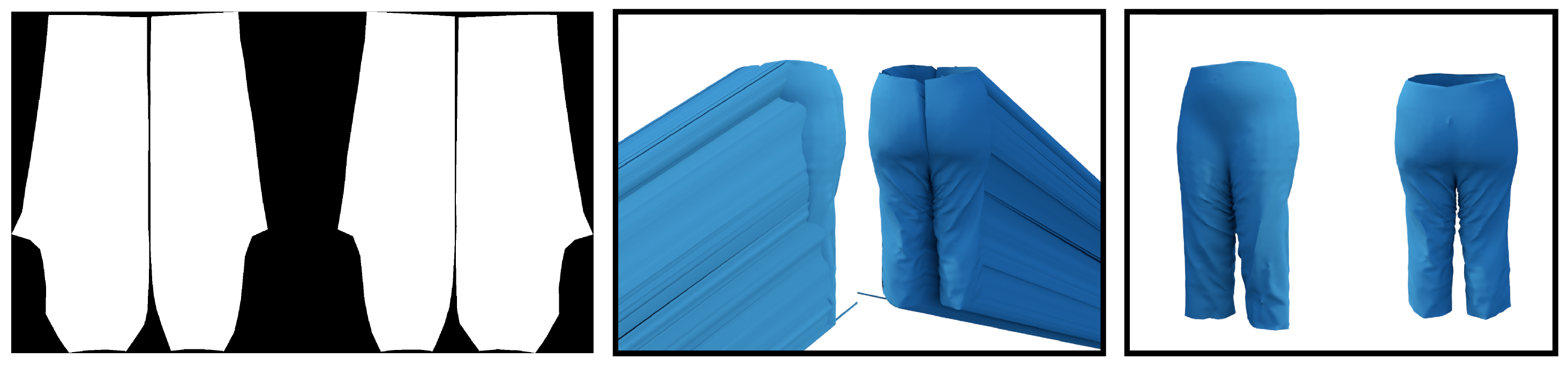} \\
	\begin{minipage}[b]{0.3\linewidth} 
    \centering (a) 
    \end{minipage}
    \begin{minipage}[b]{0.4\linewidth} 
    \centering (b) 
    \end{minipage}
    \begin{minipage}[b]{0.2\linewidth} 
    \centering (c) 
    \end{minipage}
	\caption{\small Example of pants (irregular mesh) reconstructed from its 2D image representation. We can see realistic results (c) with our proposed nearest padding method. \clr{(a) 2D pattern. (b) 3D model reconstructed from an unpadded geometry image (front and back). (c) 3D model reconstructed from padded geometry image (front and back).} 
	}
	\label{fig:pants_mesh2image}
\end{figure}

\begin{figure}[H]
	\centering
	\includegraphics[width=\linewidth]{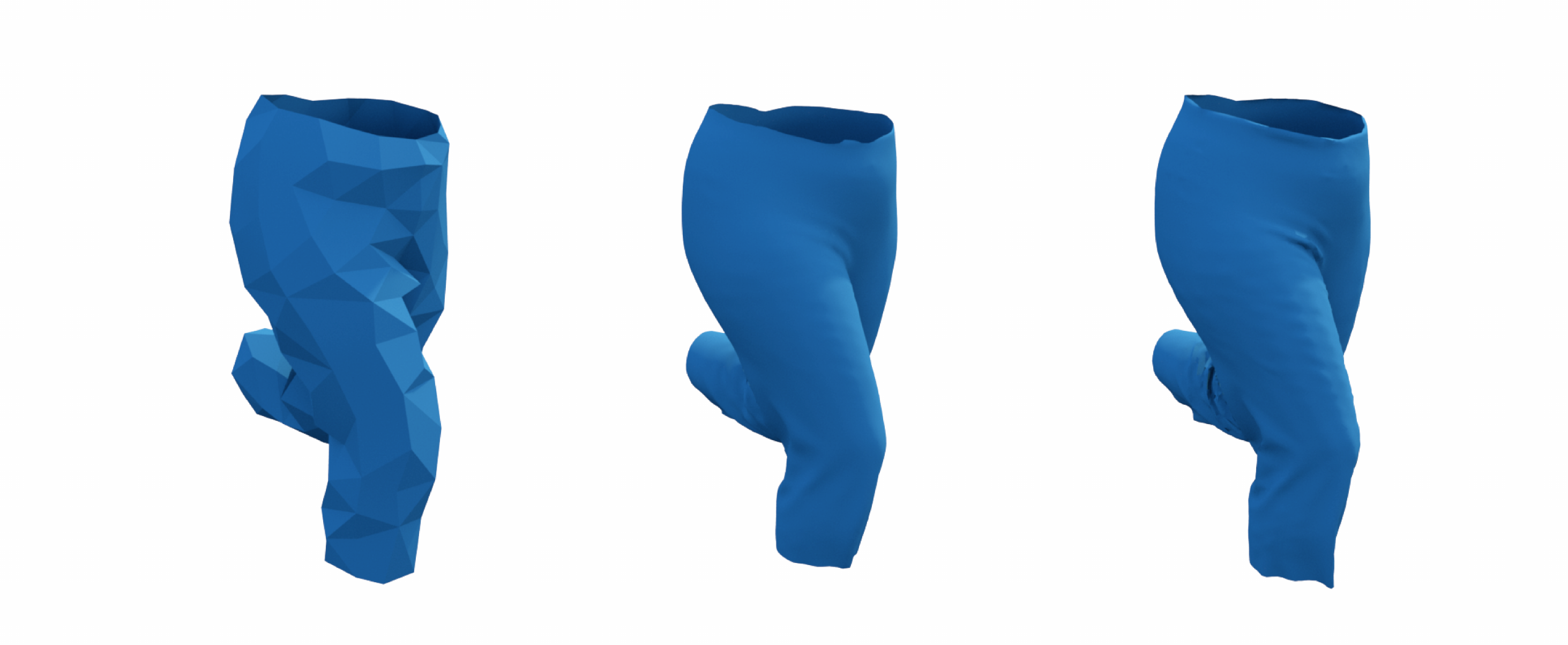} \\
	\begin{minipage}[b]{0.3\linewidth} 
    \centering (a) 
    \end{minipage}
    \begin{minipage}[b]{0.3\linewidth} 
    \centering (b) 
    \end{minipage}
    \begin{minipage}[b]{0.3\linewidth} 
    \centering (c) 
    \end{minipage}
	\caption{\small \clr{One frame result of pants. Our method can reconstruct middle-scale wrinkles. (a) Input low-resolution mesh. (b) Our reconstructed mesh. (c) Ground truth.} 
	}
	\label{fig:pants}
\end{figure}

We also evaluate our method on irregular meshes. We take pants as an example, since they are not topological a disk \clr{and} thus cannot be parametrized \clr{into} one plane. Fig. \ref{fig:pants_mesh2image} and Fig. \ref{fig:pants} show our method for pants. We follow the general idea of geometry images \cite{gu2002geometry} to cut and open the mesh into patches. The process of unfolding 3D cloth models into 2D patches is especially natural since 3D garments usually have corresponding 2D patterns in garment design. Fig. \ref{fig:pants_mesh2image} (a) shows the four 2D patterns consisted of the pants, and Fig. \ref{fig:pants_mesh2image} (b) shows the reconstructed mesh from \ref{fig:pants_mesh2image} (a) with clear artifacts at the seam lines. Our proposed padding algorithm in \clr{Subsection} \ref{sec:representation} fills the zero pixels for learning and reconstruction. \clr{In} Fig. \ref{fig:pants_mesh2image} (c), we can see the well reconstructed mesh from padded geometry images. Fig. \ref{fig:pants} shows a frame result of synthesized pants with our method, compared with input coarse mesh and the ground truth. It can be seen that our method \clr{reconstructs} middle-scale wrinkles.
}
\begin{figure*}[!htb]
\includegraphics[width=\linewidth]{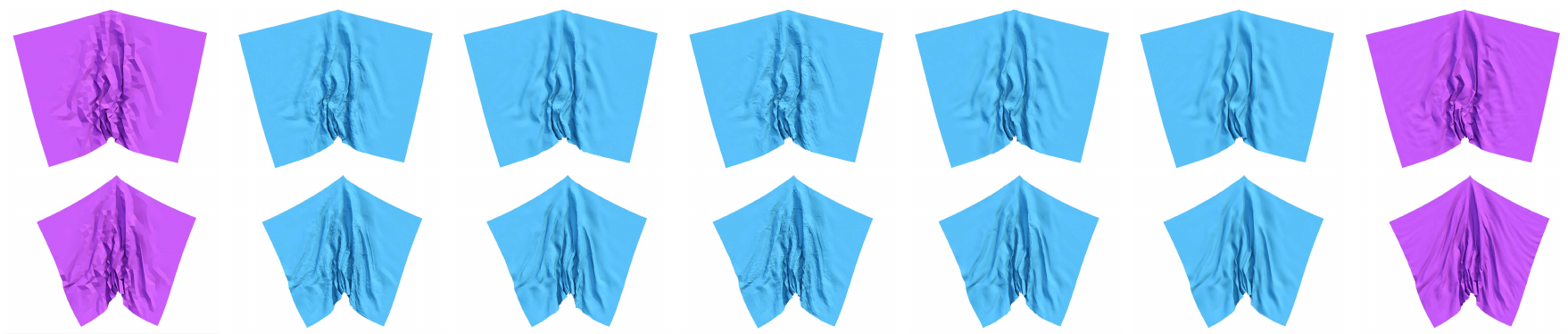} \\ 
\begin{minipage}[b]{0.138\linewidth} 
\centering (a)
\end{minipage}
\begin{minipage}[b]{0.138\linewidth} 
\centering (b)
\end{minipage} 
\begin{minipage}[b]{0.138\linewidth} 
\centering (c)
\end{minipage} 
\begin{minipage}[b]{0.138\linewidth} 
\centering (d)
\end{minipage} 
\begin{minipage}[b]{0.138\linewidth} 
\centering (e)
\end{minipage} 
\begin{minipage}[b]{0.138\linewidth} 
\centering (f)
\end{minipage} 
\begin{minipage}[b]{0.138\linewidth} 
\centering (g)
\end{minipage} 
\\
	\caption{\small \clr{ Evaluation of each loss term for our MFSR model. (a) Input LR mesh. (b) Results with  $\mathcal{L}_{d}$. (c) Results with $\mathcal{L}_{d+n}$. (d) Results with $\mathcal{L}_{d+v}$. (e) Results with  $\mathcal{L}_{d+n+v}$. (f) Our results. (g) Ground truth.
	}}
	\label{fig:Loss_Function_Comparison}
\end{figure*} 

\subsection{Ablation Study}
Next, we study the effect of different components of our proposed network, including loss function and network architecture.

\textbf{Loss function.}
To demonstrate the effectiveness of our proposed loss functions, we conduct the experiments with different loss combinations on three datasets, \clr{i.e.,} DRAPING, HITTING, and SKIRT, respectively.
The training and testing datasets are selected as mentioned in \clr{Subsection} \ref{sec:dataset}. 
We use the displacement loss as the baseline and progressively add the remaining loss terms of our \cll{MFSR}, to obtain the comparative results. 

\cll{Table~\ref{table:loss_error} reports quantitative evaluations of our proposed loss functions.
In several settings of loss functions, we compare the PSNR between generated displacement images and the ground truth, and the VMSE between synthesized meshes and the ground truth.} 
Red text indicates the best performance and the blue text indicates the second-best. 
The result shows that our algorithm has either \clr{the best or second-best performance} through combining all loss terms in a multi-task learning framework.
Notice that without the constraints of velocity and kinematics-based loss, the results with $\mathcal{L}_{n}$ gain lower PSNR \cll{and higher VMSE} although it encourages wrinkle generation in SR results.
\begin{table*}
	\caption{Comparison of Pixel-wise and Vertex-wise Error Values (PSNR/VMSE) Using Different Loss Terms on the Tablecloth (DRAPING and HITTING Separately) and SKIRT Datasets }
	\label{table:loss_error}
	\centering 
	\begin{tabular}{cccccc}
		\hline
		\multirow{2}{*}{Dataset} & $\mathcal{L}_{d}$ &  $\mathcal{L}_{d+n}$ & $\mathcal{L}_{d+v}$ & 
		$\mathcal{L}_{d+n+v}$ & $\mathcal{L}_{all}$ \\ \cline{2-6}
		& PSNR$\uparrow$/VMSE$\downarrow$  & PSNR$\uparrow$/VMSE$\downarrow$ & PSNR$\uparrow$/VMSE$\downarrow$ & PSNR$\uparrow$/VMSE$\downarrow$ & PSNR$\uparrow$/VMSE$\downarrow$ \\ \hline
		DRAPING  &  67.90/9.44e-5 & 62.83/1.94e-4 & \textbf{\color{blue} 67.92/9.43e-5} & 63.47/1.84e-4 & \textbf{\color{red} 68.68/7.26e-5}       \\  \hline
		HITTING & 67.11/9.37e-5 & 68.23/7.67e-5 & \textbf{\color{red} 71.41/5.48e-5} & 69.75/6.26e-5 & \textbf{\color{blue} 71.26/5.61e-5}
		\\  \hline
		SKIRT & 62.48/1.92e-4 & 60.76/2.72e-4 &\textbf{\color{blue} 62.48/1.91e-4} & 61.31/3.02e-4 &\textbf{\color{red} 62.88/1.82e-4}
		\\ \hline
	\end{tabular}
	\begin{tablenotes}
		\item \footnotesize Note: The number in red/blue indicates the best/second-best performance. ``$\uparrow$'' means higher is better. ``$\downarrow$'' means lower is better.
	\end{tablenotes}
\end{table*}

In Fig. \ref{fig:Loss_Function_Comparison}, we show visual results from the DRAPING dataset, to evaluate the performance of different loss combinations.
The baseline model (see Fig.\ref{fig:Loss_Function_Comparison}(a)) only including displacement loss generates cloth meshes with too many unrealistic small wrinkles and inconsistent thin lines.
Directly combining velocity loss with the baseline (see Fig.\ref{fig:Loss_Function_Comparison}(c)) is not able to solve this problem, contrarily introducing some unexpected horizontal short lines.  
The results including normal loss also suffer from uneven lines and sharp triangles on the folders and boundaries (\clr{i.e.}, several buckling triangles near the handler and center wrinkle lines on the top of \clr{Fig.\ref{fig:Loss_Function_Comparison}(b) and \ref{fig:Loss_Function_Comparison}(d))}.
Finally, the results with all our proposed loss functions (see Fig.\ref{fig:Loss_Function_Comparison}(e)) are visually very close to the ground truth (see Fig.\ref{fig:Loss_Function_Comparison}(f)) with the realistic and consistent wrinkles and folders, which verifies the effectiveness of our proposed four loss functions.

\begin{figure}[H]
	\centering
 \includegraphics[width=\linewidth]{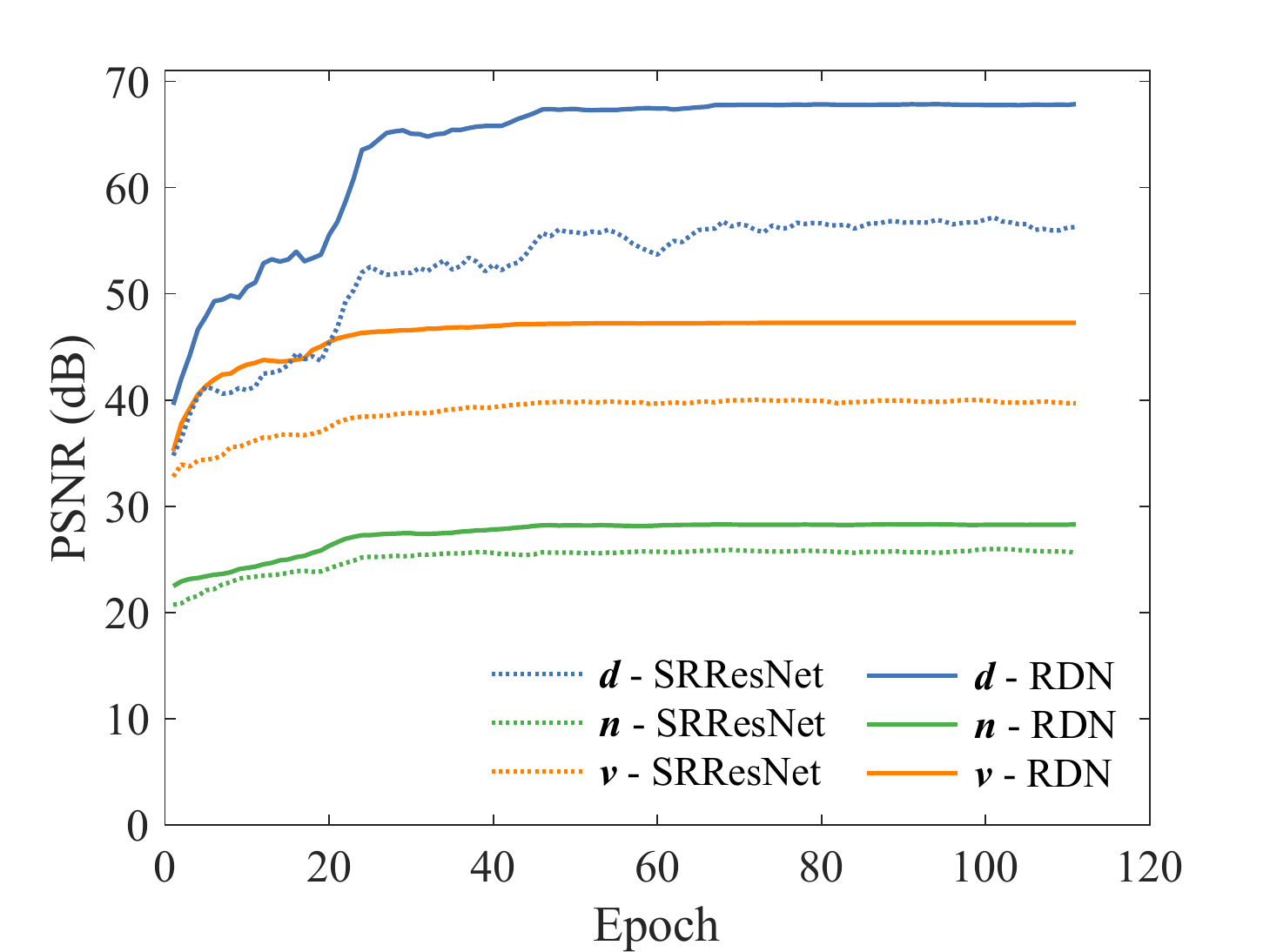} 
	\caption{\small \cl{Convergence analysis of the displacement $\bm{d}$, the normal $\bm{n}$ and the velocity $\bm{v}$ with RDN  \cite{zhang2018residual} and SRResNet \cite{ledig2017photo}. PSNR between the inference of super-resolution networks and the ground truth is used for this quantitative evaluation, where RDN achieves better performance.}
	}
	\label{fig:curve_rdn_resnet}
\end{figure} 
\textbf{Network architecture.}
To further investigate the performance of different networks (SRResNet and RDN), we conduct experiments on the DRAPING dataset.
In particular, we validate our cloth animation results on randomly selected 800 pairs of LR/HR meshes from the DRAPING dataset, 
which are excluded from the training set, and cover different complex hanging motions in pendulum movement. 
In Fig. \ref{fig:curve_rdn_resnet}, we depict the convergence curves of three different features in the above validation dataset.
The convergence curves show that RDN achieves better performance than SRResNet, and further stabilizes the training process in all three features.
The improved performance and stabilization are benefited from contiguous memory, local residual learning and global feature fusion in RDN.
In SRResNet, local convolutional layers do not have direct access to the subsequent layers, \clr{thereby it neglects to fully use the information of each convolutional layer.} 
As a result, RDN achieves better performance than SRResNet.

\section{\clr{Conclusions}}
\clr{
This paper proposed a novel multi-feature super-resolution network to generate winkle details from coarse simulated meshes. By using multiple features as well as corresponding losses, our system performed well in predicting mesh sequences stably and realistically.
Experimental results revealed that our system can effectively synthesize dynamic wrinkles maintaining temporal consistency and achieve a high frame rate in various scenes (such as cloth and outfits). Meanwhile, the proposed synchronized simulation contributed to construct dataset of paired 3D meshes. These aligned coarse and fine meshes can also be used for other applications such as 3D shape matching of incompatible shape structures.
}

\clr{Nevertheless, several limitations remain open for the future work.} 
In our work, the training data is the paired LR/HR meshes generated by a synchronized simulation.
While tracking the LR cloth, the HR cloth cannot show dynamic properties of a full simulation.
We would like to address this limitation by imposing unsupervised learning \cl{or cycle/dual generative adversarial networks to learn a mapping between the high-resolution meshes and low-resolution meshes in the future. In addition, the dataset should be further expanded including more scenes, motion sequences, and garment shapes to create more diverse results. Our work is not independent of physics based simulation, but is an acceleration one. Thus the estimated wrinkles from networks are related to the materials setting in physics-based simulation phase. 
It could be an interesting future direction to generalize our method to diverse materials \clr{and thus} generate different types of wrinkles.}


\vspace{2mm}


\vspace{3mm}

\footnotesize
\itemsep=-3pt plus.2pt minus.2pt
\baselineskip=14pt plus.2pt minus.2pt
\bibliographystyle{myJCST}
\bibliography{reference_new} 
 
\vspace{5mm}

\clr{
\begin{spacing}{1.5}
\noindent\parbox{8.3cm}{\parpic{\includegraphics[width=1in,height=1.25in,clip,keepaspectratio]{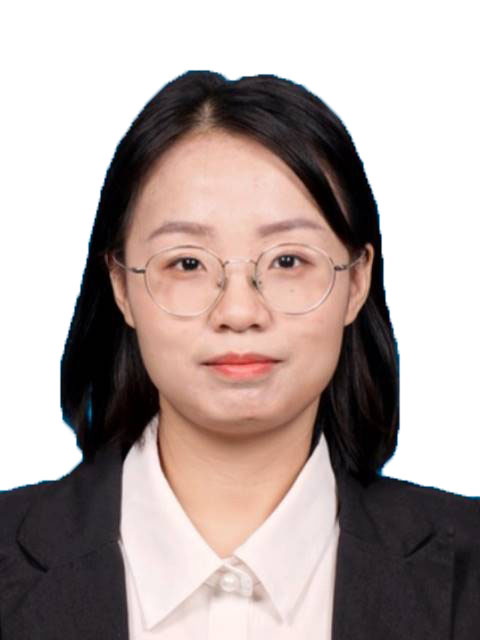}}{\small\quad {\bf Lan Chen} received her bachelor's degree in mathematics from China University of Petroleum, Beijing, China, in 2016. She is currently a Ph.D. candidate at Institute of Automation, Chinese Academy of Sciences, Beijing, China. Her research interests include computer graphics, geometry processing and image processing, particularly synthesis of cloth animation.}\\[1mm]}

\noindent\parbox{8.3cm}{\parpic{\includegraphics[width=1in,height=1.25in,clip,keepaspectratio]{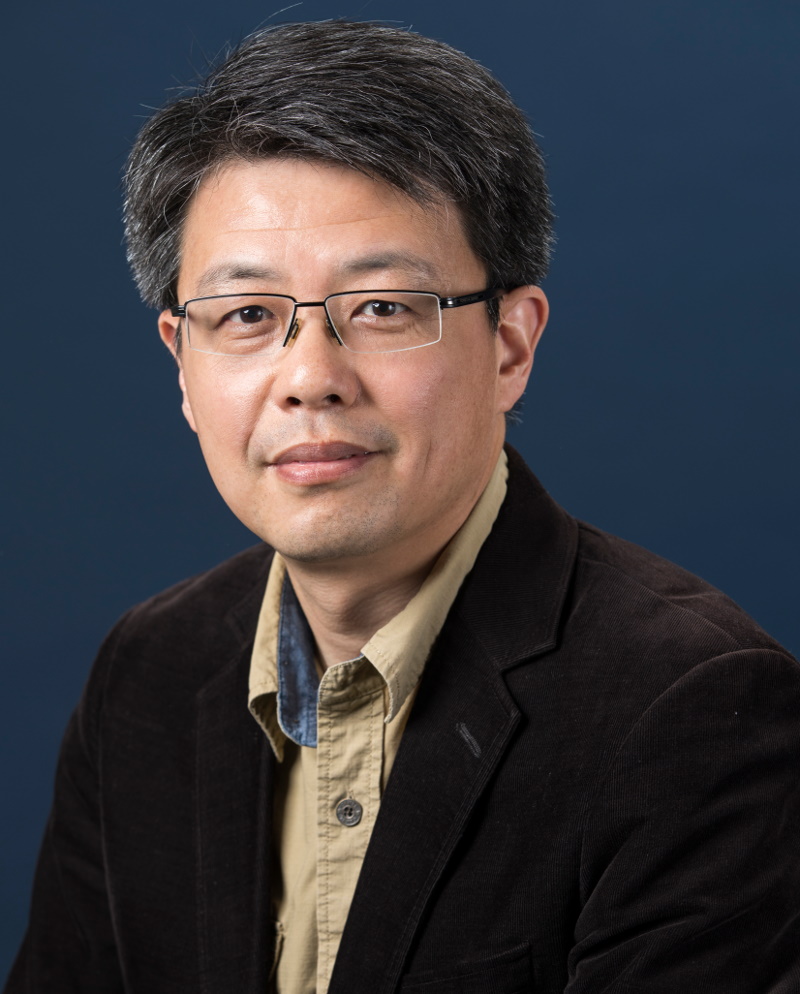}}{\small\quad {\bf Juntao Ye} was awarded his bachelor's degree from Harbin Engineering University, Heilongjiang Province, China, in 1994, and the M.S. degree from Institute of Computational Mathematics and Sci/Eng Computing, Chinese Academy of Sciences, Beijing, China, in 2000, and the PhD degree in Computer Science from University of Western Ontario, Canada, in 2005. He is currently an associate professor at Institute of Automation, Chinese Academy of Sciences, Beijing, China. His research interests include computer graphics and image processing, particularly simulation of cloth and fluid.}\\[1mm]}

\noindent\parbox{8.3cm}{\parpic{\includegraphics[width=1in,height=1.25in,clip,keepaspectratio]{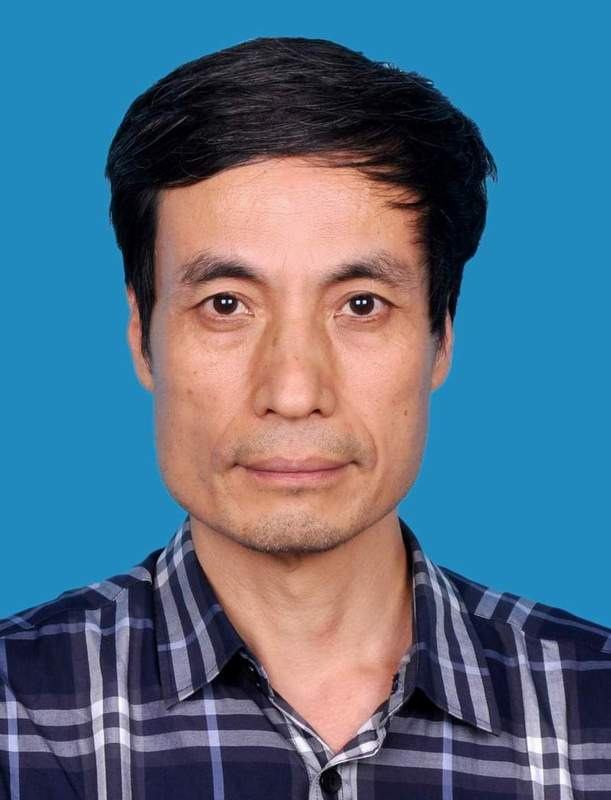}}{\small\quad {\bf Xiaopeng Zhang} received the PhD degree in computer science from Institute of Software, Chinese Academic of Sciences, Beijing, China, in 1999. He is a professor in National Laboratory of Pattern Recognition at Institute of Automation, Chinese Academy of Sciences, Beijing, China. He received the National Scientific and Technological Progress Prize (second class) in 2004 and the Chinese Award of Excellent Patents in 2012. His main research interests include computer graphics and computer vision.}\\[1mm]}
\end{spacing}
}

\label{last-page}
\end{multicols}
\label{last-page}
\end{document}